\documentclass{aa}  

\usepackage{graphicx}
\graphicspath{{Figures/}}

\usepackage{txfonts}
\usepackage{lipsum}
\usepackage{subcaption}         
\usepackage{lscape}             
\usepackage{placeins}           

\usepackage[table]{xcolor}  
\usepackage{multirow}      
\usepackage{arydshln}      
                                
\usepackage[colorlinks=true,
            linkcolor=blue,
            citecolor=blue,
            urlcolor=blue]{hyperref} 

\makeatletter
\DeclareRobustCommand{\citealias}[2]{%
  \hyperlink{cite.#1}{#2}%
}
\makeatother

\usepackage{natbib}

\begin{document}

   \title{Search for magnetic fields in seven slowly rotating A stars}


%

   \author{P. de Frutos-Rull\inst{1} \fnmsep\thanks{Corresponding author: pablo.de-frutos-rull@obspm.fr}
        \and C. Neiner\inst{1}
        \and J. Labadie-Bartz\inst{2,1}
        \and F. Royer\inst{1}
        \and R. Monier\inst{1}
        \and K. Thomson-Paressant\inst{3,1}
        }

   \institute{LIRA, Observatoire de Paris, Universit\'e PSL, CNRS, Sorbonne Universit\'e, Universit\'e Paris Cit\'e, CY Cergy Paris Universit\'e, 92190 Meudon, France
   \and DTU Space, Technical University of Denmark, Elektrovej 327, Kgs., Lyngby 2800, Denmark
   \and School of Mathematics, Statistics and Physics, Newcastle University, Newcastle upon Tyne, NE1 7RU, UK}

   \date{Received 28 May 2026 / Accepted 16 July 2026}

 
  \abstract
   {A small fraction of A-type stars host strong fossil magnetic fields. Identifying them among the broader A-type population relies on indirect indicators such as chemical peculiarities, slow rotation, and rotational modulation, which are also shared, to varying degrees, by non-magnetic stars. Assessing the reliability of these indicators is essential for future surveys.}
   {We aim to magnetically characterise a sample of slowly rotating, chemically peculiar A-type stars selected through two complementary strategies, and to investigate the connections between selection criteria, rotation, and  magnetic fields.} 
   {We obtained high-resolution spectropolarimetric observations of seven slowly rotating, chemically peculiar A-type stars. Three were selected from abundance analysis combined with potential intrinsically slow rotation, and four from a spectral depression at 5200~\AA\ combined with rotational modulation in TESS photometry. We applied the least squares deconvolution method, computed false alarm probabilities, and measured longitudinal field strengths for the detections. For the non-detections, we derived upper limits on the polar field strength and evaluated Zahn and Spruit's critical fields.}
{We detect magnetic fields in all four stars selected via the spectral depression and photometric criterion, with longitudinal field strengths of $\mathord{\sim}145$--$2900$ G, while none of the three stars selected via the abundance and slow rotation criterion shows a Zeeman signature. The combined upper limits for two non-detections lie below or near the magnetic-desert boundary, but only one is also below the critical fields required for rigid rotation. The non-detections are most naturally interpreted as Am stars. Among the detections, HD~63843 stands out as a magnetic $\delta$~Scuti pulsator, promising for magneto-asteroseismology.}
   {Slow rotation combined with chemical peculiarity alone is not sufficient to reliably trace large-scale magnetic fields, whereas the addition of the 5200~\AA\ depression and stable rotational modulation in space photometry appears to select magnetic A-type stars with high efficiency, providing a practical guideline for future spectropolarimetric surveys.}


   \maketitle

\nolinenumbers
\section{Introduction}
Magnetic fields are a fundamental ingredient in the physics of stars across the Hertzsprung–Russell diagram. A small fraction of hot stars (about 10\% of all OBA stars) host strong, stable, large-scale magnetic fields 
which can most often be modelled by a simple oblique dipole configuration. These stable fields are thought to be of fossil origin \citep{NeinerMathis2015}, i.e. inherited from the molecular cloud from which the star was formed, although some could also arise from a merger process \citep{schneider2024}.

Such large-scale magnetic fields significantly affect the stellar structure and surface properties. In radiative envelopes, magnetic stresses can efficiently redistribute angular momentum, promoting rigid rotation while, at the same time, being able to suppress macroscopic mixing processes \citep{Fuller2019,Briquet2012}. This creates the conditions required for radiative levitation and gravitational settling to operate, giving rise to chemical peculiarities and surface spots observed through their spectroscopic and photometric footprints \citep{Michaud1970, alecian2023}. These long-lived and highly stable spots provide clear empirical evidence that the magnetic field arrests differential rotation at the surface, and constitute one of the clearest observational manifestations of the close links between fossil magnetism, chemical peculiarities, and rotation in hot stars.

Among the chemically peculiar (CP) star classes introduced by \cite{Preston1974}, the CP1 (Am) and CP2 (Ap) stars are two important groups in the A-type regime, and they provide a useful contrast for studying the role of magnetism in stellar structure and evolution. Am stars, also known as metallic-line A-type stars, exhibit an apparent underabundance of Ca or Sc, along with an overabundance of Fe-group and heavier elements; nearly all heavy elements (with a few exceptions) in Am stars are enhanced in the stellar photosphere. By contrast, in Ap stars only specific elements (such as Si, Sr, Eu, Cr, and rare-earth elements) are greatly enriched in abundance. In Ap stars, these abundance anomalies are typically organised into stable surface inhomogeneities associated with the magnetic field structure, leading to clear, periodic rotational modulation of spectral lines and flux, offering observable indirect manifestations of stable magnetic fields in spectroscopy and photometry. Photometric variability has also been reported in Am stars \citep{Balona2015}, but the detected rotational signals are generally low in amplitude and may vary in amplitude or frequency over time, in contrast to the more stable modulation typical of Ap stars.

The magnetic properties of these two classes are likewise contrasted. Ap stars host large-scale fields with polar strengths of the order of a few kG at the surface \citep{Shultz2019}, with a  lower bound of $\mathord{\sim}300$ G, the limit of the so-called magnetic desert that separates Ap stars from the rest of the A-type population \citep{Auriere2007,Lignieres2014}. Am stars, by contrast, are not known to host strong large-scale fields; the few magnetic detections reported in Am stars correspond to ultra-weak fields at the sub-gauss level \citep{Blazere2018}, below the magnetic desert and orders of magnitude weaker than the fields characteristic of Ap stars.

Within this framework, rotation plays a particularly important role. As shown by \cite{Abt1995}, both Ap and Am stars are, on average, significantly slower rotators than their non-magnetic counterparts. The nature of this slow rotation in Ap stars has long remained an open question \citep{Stepien1998}. While in some cases it may reflect magnetic braking during early evolutionary phases \citep{Shultz2019}, the relatively weak winds of A-type stars limit the efficiency of this mechanism.  Consequently, slow rotation may also preserve information about initial conditions at birth, offering a window into pre-main-sequence evolution and the origin of fossil magnetic fields \citep{Kochukhov2006}. 
The fact that slow rotation is generally shared by both Ap and Am stars (although some Ap stars are fast rotators with periods of $\lesssim\!1$ day) also means that, taken on its own, it does not unambiguously identify the magnetic class. Understanding the interplay between rotation, chemical peculiarities, and magnetism in A-type stars could therefore provide important insight into this domain of stellar physics.

Progress in this area is especially timely in the context of the upcoming PLATO mission \citep{PLATO}, which is expected to deliver high-precision light curves for large samples of intermediate-mass stars and thus enable their asteroseismic characterisation. Combined with spectropolarimetric measurements, this will open the door to magneto-asteroseismology, the joint probing of internal structure and magnetic fields \citep[e.g.][]{Neiner2021}. Realising this potential requires improved observational constraints on the incidence and properties of magnetic fields in well-characterised stellar samples.

In this work, we focus on a set of CP A stars selected for their low projected rotational velocities and indirect indicators of magnetism, drawn from two complementary selection strategies. By conducting a high-resolution spectropolarimetric analysis of these targets, we aim to characterise their magnetic properties and assess the occurrence of large-scale magnetic fields in slowly rotating A stars. The resulting detections and non-detections, combined with a re-examination of the available abundance information and complementary photometric diagnostics, provide new insights into the magnetic and chemical nature of these objects. More broadly, this study contributes to the observational groundwork required for future magneto-asteroseismic investigations enabled by forthcoming space missions.

\section{Selection of targets} \label{Sect: selection of targets}

The sample analysed in this study comprises seven CP stars with relatively low projected rotational velocities ($v\sin i \lesssim\!45 \,\text{km}\,\text{s}^{-1}$), identified as promising candidates for hosting magnetic fields based on indirect indicators such as chemical peculiarities, rotational modulation, and potential intrinsic slow rotation. The targets were selected according to two different criteria, each described in a dedicated subsection below. 

\begin{table*}
\captionsetup{justification=raggedright,singlelinecheck=false}
\caption{Target properties and parameters adopted for the initial VALD3 line masks.}
\label{Tab:target_properties}
\centering
\begin{tabular}{lllllll}
\hline
\noalign{\smallskip}
Target & TIC & Spectral type & $V$ & $v\sin i$ & $T_{\,\rm eff}^{\, \rm mask}$ & $\log g^{\, \rm mask}$ \\
& & & (mag) & (km\,s$^{-1}$) & (K) & \\
\noalign{\smallskip}
\hline\hline
\noalign{\smallskip}
HD~65900  & 452843608 & A1V, CP          & 5.64  & $36.4 \pm 1.5$ & 9250 & 4.0 \\
HD~154228 & 142696134 & A1V, CP          & 5.908 & $45.2 \pm 1.2$ & 9250 & 4.0 \\
HD~158716 & 400150498 & A1V, CP          & 6.48  & $6.4 \pm 0.8$  & 9250 & 4.0 \\
\noalign{\smallskip}
\cdashline{1-7}
\noalign{\smallskip}
HD~63843     & 35884762  & A2IVSrCrEu       & 10.25 & $7 \pm 2 \: ^{*}$ & 7750 & 3.5 \\
HD~266267    & 235391838 & A7VSrEu          & 10.00 & $5\pm3$           & 8750 & 4.0 \\
BD~+01~1920  & 271375640 & A1--A7SrCrEuSi   & 10.00 & $5 \pm 1$         & 8750 & 4.0 \\
BD~+08~2211  & 312111544 & A3--A9SrCrEu     & 9.20  & $6\pm1$           & 7250 & 4.0 \\
\noalign{\smallskip}
\hline
\end{tabular}

\tablefoot{
Columns list the target identifiers, TIC number, spectral type, $V$ magnitude, projected rotational velocity, and the effective temperature and surface gravity of the template masks. Stars above the dashed line were selected from \citealias{Royer2014}{R14}, and their spectral types and Fourier-method determined $v\sin{i}$ values are from \citealias{Royer2014}{R14}. Those below the dashed line are from \citealias{Thomson-Paressant2024}{TP24}, with spectral types from \cite{Hummerich2020} and Fourier-method determined $v\sin{i}$ values from \citealias{Thomson-Paressant2024}{TP24}. Note that a new calculation of the $v\sin{i}$ value for HD~63843 was performed with the Fourier method for this work, yielding a value of $15.36 \pm 0.3$ km\,s$^{-1}$, in closer agreement with the features observed in Figures~\ref{fig:LSD_plots} and~\ref{fig:Integration_limits} than the previous value marked with an asterisk.}

\end{table*}

\subsection{Slow rotation} \label{Sect: SlowRot}
The first subset was drawn from the study of \cite{Royer2014}, hereafter \citealias{Royer2014}{R14}, who investigated A0–A1 stars with low projected rotational velocities ($v\sin i \leq 65$ km\,s$^{-1}$) using high-resolution and high signal-to-noise spectroscopy with the SOPHIE and \'ELODIE spectrographs at Observatoire de Haute-Provence to identify CP stars and spectroscopic binaries. Based on a detailed abundance analysis and a hierarchical classification of chemical patterns, they isolated a subsample of CP A stars, without assigning specific CP subtypes such as CP1 (Am) or CP2 (Ap).  Their statistical analysis of the rotational velocity distribution, corrected for projection effects under the assumption of randomly oriented rotation axes, indicated that CP stars in their sample are unlikely to be fast rotators seen pole-on. 
The targets selected with this criterion for this study are HD~65900, HD~154228, and HD~158716.

The detailed chemical analysis of these three stars by \citealias{Royer2014}{R14} revealed iron-peak and heavy-element overabundances together with a Sc deficit in all three stars (strong in HD~158716, clear in HD~154228, milder in HD~65900) and a significant Ca deficit in HD~158716 only, with HD~154228 and HD~65900 showing near-solar Ca abundances. Si is overabundant in the three stars. The \citealias{Royer2014}{R14} abundance set does not include exhaustive rare-earth diagnostics, and a definitive assignment to a specific CP subtype is therefore non-trivial from these data alone. The combination of slow rotation and chemical peculiarities in these stars makes them well-suited targets for spectropolarimetric characterisation, providing complementary diagnostic information on their magnetic and chemical nature and motivating their inclusion in the present study.

\subsection{Rotational modulation in the TESS light curve}
The second subset was drawn from the selection criterion introduced by \cite{Thomson-Paressant2024}, hereafter \citealias{Thomson-Paressant2024}{TP24}, which combines photometric and spectroscopic indirect indicators of magnetism, namely the presence of a spectral depression at 5200~\AA\ first catalogued by \cite{Hummerich2020} and clear rotational modulation detected in the TESS light curves from the analysis presented by \cite{Labadie-Bartz2023}. This approach is designed to identify CP stars that exhibit photometric variability due to the rotational modulation of surface abundance inhomogeneities, a hallmark of magnetic CP stars. From this sample, we selected CP2 objects with some of the lowest projected rotational velocities, namely HD~63843, HD~266267, BD~+01~1920, and BD~+08~2211 (all with $v \sin i \lesssim 15~{\rm km~s^{-1}}$), as prime candidates for studying the interplay between slow rotation and magnetism. Although \citealias{Thomson-Paressant2024}{TP24} already report magnetic signatures for these stars, they are included here to extend and refine the spectropolarimetric characterisation of these slow rotators, building on and complementing the methodology developed therein.

\section{Spectropolarimetric observations}
All of the stars analysed in this work were observed using the Echelle SpectroPolarimetric Device for the Observation of Stars \citep[ESPaDOnS;][]{Donati_ESPaDOnS} mounted on the Canada–France–Hawaii Telescope (CFHT) at Mauna Kea Observatory. The spectropolarimetric data from ESPaDOnS were acquired in circular polarisation mode during observing runs in January 2024 and April 2025. Each polarimetric measurement consists of a sequence of four sub-exposures, which are combined to yield one Stokes~$I$ and one Stokes~$V$ spectrum. A summary of the observed targets, along with the details of their observations, is presented in Tables \ref{Tab:target_properties} and \ref{Tab:observations}. 

For each target, the exposure time was chosen using an empirical ESPaDOnS S/N--exposure-time relation calibrated from archival observations, with the aim of reaching sensitivity to dipolar magnetic fields with a polar field value down to 100 G. The required S/N depends on the apparent magnitude, spectral type, and $v\sin{i}$: the spectral type determines the approximate number and strength of lines entering the LSD mask, while low $v\sin{i}$ values make weaker Zeeman signatures more easily detectable. Given the field strengths typical of Ap stars, the sensitivity threshold adopted here is expected to be adequate for detecting magnetic fields in the stars comprising this sample.

In addition to the ESPaDOnS data, one target (HD~158716) had also been observed earlier with the Narval spectropolarimeter \citep{Auriere_Narval} installed on the Telescope Bernard Lyot (TBL) at the Pic du Midi Observatory. These archival data, from July 2014, were incorporated into the present analysis.

\section{Spectropolarimetric analysis}

The ESPaDOnS and Narval spectropolarimetric data were processed using the \textsc{LibreEsprit} reduction package \citep{Donati1997}, together with the \textsc{Upena} pipeline \citep{Martioli2011}. The continuum normalisation was performed independently using the \textsc{SpeNT} software \citep{Martin2018}. Line masks required for the subsequent analysis were generated from the Vienna Atomic Line Database \citep[VALD3;][]{VALD3}. 
For each target, the initial template mask was computed using approximate atmospheric parameters selected to match the star’s spectral type and observed spectrum. These parameters, listed in Table~\ref{Tab:target_properties}, were used only to define the starting VALD3 line list. These initial template masks were then refined by excluding spectral regions affected by hydrogen lines, telluric absorption, and other features deemed unsuitable for line combination. The depths of the retained lines were then adjusted to reproduce the observed line strengths in the stellar spectra, following the procedure outlined by \cite{Grunhut2017}. 

The spectropolarimetric analysis of CP stars requires particular care in the construction of line masks. In addition to accurately adjusting line depths, which often deviate significantly from those expected for normal stars with similar parameters, it is frequently necessary to add spectral lines present in the observed spectra but absent from the initial template masks. For the stars in our sample, the standard VALD3-based template masks proved insufficient due to pronounced chemical peculiarities. In these cases, additional lines were identified directly from the observed spectra and incorporated into the masks using information from alternative VALD3 template masks computed with different effective temperatures or surface gravities. For targets in common with the sample analysed by \citealias{Thomson-Paressant2024}{TP24}, we adopted the line masks constructed in their study as a starting point. These masks were subsequently modified to ensure consistency with our analysis.

\begin{figure*}
\centering
\includegraphics[width=18cm]{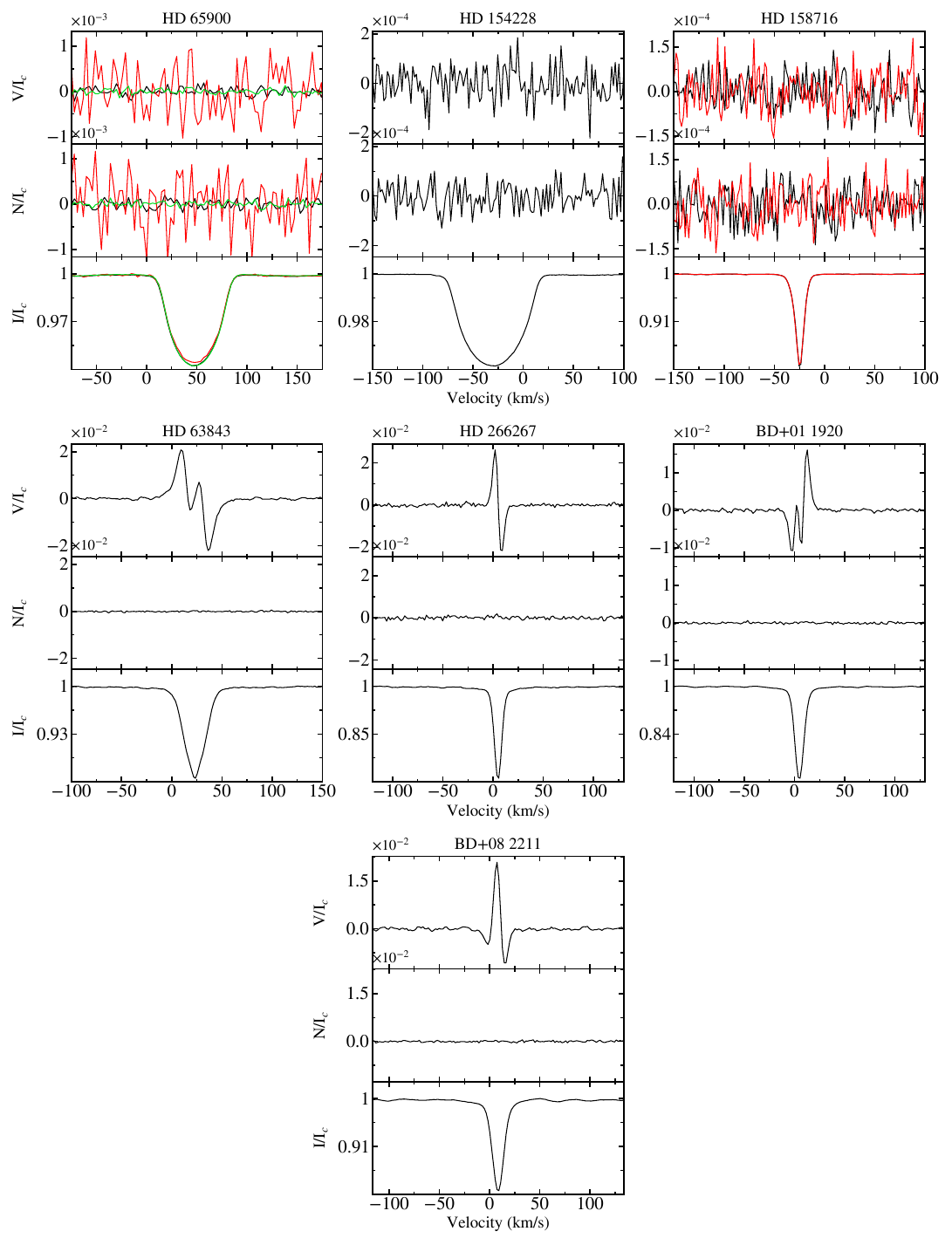}
\caption{LSD Stokes~$V$ (top panel), diagnostic~$N$ (middle panel), and Stokes~$I$ (bottom panel) profiles for all the targets, shown in 250 km/s windows around the centre of the profiles. When multiple observations are available, each is shown in a different colour.}
\label{fig:LSD_plots}
\end{figure*}

The extraction of mean line profiles was carried out using the least squares deconvolution (LSD) technique \citep{Donati1997}. This approach combines numerous individual spectral lines, weighted according to their depth, Landé factor, and central wavelength, to produce Stokes~$I$ and $V$ profiles with high signal-to-noise ratio. In addition to these profiles, the LSD procedure yields a diagnostic null $N$ profile, which serves as a check for spurious signatures and allows the robustness of any detected magnetic signal to be assessed. The plots for the LSD profiles for all the targets in this work can be found in Figure \ref{fig:LSD_plots}. 

\section{Magnetic characterisation}
In order to quantitatively assess whether a magnetic signature was detected in a given target, we computed the false alarm probability (FAP). After LSD deconvolution, $\chi^2$ statistics were evaluated both inside and outside the stellar line region. Comparing the measured Stokes~$V$ profile with the noise characteristics yields a certain probability that the magnetic signature in the Stokes~$V$ profile is produced by noise; this is the FAP. Following \cite{Donati1992}, values below $10^{-5}$ indicate a definite magnetic detection (DD, with a detection probability of 99.999\%), values between $10^{-5}$ and $10^{-3}$ correspond to marginal detections (MD), and values above $10^{-3}$ imply no significant signal (ND). 

The resulting FAP values for all targets in our sample are presented in the third column of Tables \ref{Tab:DDs_characterisation} and \ref{Tab:NDs_characterisation}. From these results, we conclude that none of the three stars selected from \citealias{Royer2014}{R14} show evidence of a magnetic Zeeman signature, whereas all stars selected from \citealias{Thomson-Paressant2024}{TP24} exhibit clear magnetic detections. In the following subsections, we provide a magnetic characterisation of the targets, treating the cases of magnetic detections and non-detections separately, as they require different approaches.

\subsection{Longitudinal field calculation for magnetic detections}
For targets exhibiting definite magnetic signatures, we calculated the longitudinal magnetic field $B_l$ by integrating the LSD Stokes~$I$ and $V$ profiles for each star, following the centre of gravity method described by \cite{Rees1979} and \cite{Wade2000}. The resulting measurements are reported in Table \ref{Tab:DDs_characterisation}.

A key element of this calculation is the definition of the velocity interval over which the profiles are integrated. In this work, the integration region was defined around the line centroid so as to encompass the full spectral line while minimising the contribution from the surrounding continuum. The adopted velocity limits for the magnetic targets are listed in column 5 of Table \ref{Tab:DDs_characterisation} and are illustrated in Figure \ref{fig:Integration_limits}.

Table \ref{Tab:DDs_characterisation} also reports the corresponding values of $N_l$, computed in an identical manner by replacing the Stokes~$V$ profiles with the diagnostic $N$ profiles. For comparison, column 8 lists the longitudinal magnetic field strengths reported by \citealias{Thomson-Paressant2024}{TP24}. With the exception of HD~63843, our measurements are consistent within error bars with those of \citealias{Thomson-Paressant2024}{TP24}. Although the line masks used in the two studies differ slightly (ours being derived from those of \citealias{Thomson-Paressant2024}{TP24} but employing a slightly different spectral normalisation, line cleaning, and line depths), the discrepancy for HD~63843 is most likely attributable to the choice of integration range. In particular, the velocity interval adopted by \citealias{Thomson-Paressant2024}{TP24} for this star (shown in Figure \ref{fig:Integration_limits}) is significantly narrower than that used in the present analysis and truncates the wings of the stellar line, therefore likely underestimating the magnetic signal.

\begin{table*} 
\centering
\caption{False alarm probabilities, integration parameters, and longitudinal magnetic field measurements for the four stars with definite magnetic detections.}
\begin{tabular}{l c c c c c c c}
\hline
\noalign{\smallskip}
Star ID & Date & FAP & $v_{\text{centroid}}$ (km\,s$^{-1}$) &
Integration range (km\,s$^{-1}$) & $N_l$ (G) & $B_l$ (G) & $B_l^{\text{lit}}$ (G) \\
\noalign{\smallskip}
\hline\hline
\noalign{\smallskip}
HD~63843    & 19-Jan-24 & $<10^{-5}$ (DD) & 23.1 & $[-14.9$, $61.1]$ & $-8 \pm 21$ & $2901 \pm 32$ & $2191 \pm 15$ \\
HD~266267   & 16-Jan-24 & $<10^{-5}$ (DD) &  5.0 & $[-10.0$, $20.0]$ & $13 \pm 14$ & $396 \pm 14$  & $403 \pm 11$ \\
BD~+01~1920 & 08-Jan-24 & $<10^{-5}$ (DD) &  5.0 & $[-15.0$, $25.0]$ & $-4 \pm 7$  & $-417 \pm 8$  & $-416 \pm 7$ \\
BD~+08~2211 & 05-Jan-24 & $<10^{-5}$ (DD) &  8.5 & $[-12.5$, $29.5]$ & $19 \pm 8$  & $145 \pm 15$  & $152 \pm 13$ \\
\hline
\end{tabular}

\tablefoot{
Columns list the star name, the date of observation, the FAP detection status, the velocity of the centroid of the line ($v_{\text{centroid}}$), the velocity range used for integration, the longitudinal field measured from the diagnostic null profile ($N_l$), the longitudinal field in gauss measured from Stokes~$V$ ($B_l$), with uncertainties from error propagation, and the literature value $B_l^{\text{lit}}$ taken from \citealias{Thomson-Paressant2024}{TP24}.}

\label{Tab:DDs_characterisation}
\end{table*}

\begin{table*}
\captionsetup{justification=raggedright,singlelinecheck=false}
\caption{False alarm probabilities and upper limits on the polar magnetic field strength for the targets with no field detected.}
\centering
\begin{tabular}{l c c c c}
\hline
\noalign{\smallskip}
Star ID & Date & FAP & $B_{\text{pol,max}}$ (G) & $B_{\text{pol,max}}^{\text{comb}}$ (G) \\
\noalign{\smallskip}
\hline\hline
\noalign{\smallskip}

HD~65900 & 04-Jan-24 & 0.6430 (ND) & $262 \pm 52$ & \multirow{2}{*}{$98 \pm 15$} \\
         & 08-Jan-24 & 0.3919 (ND) & $137 \pm 27$  &  \\

\noalign{\smallskip}
\cdashline{1-5}[0.5pt/2pt]
\noalign{\smallskip}

HD~154228 (33 Oph) & 07-Apr-25 & 0.1510 (ND) & $210 \pm 42$ &  \\

\noalign{\smallskip}
\cdashline{1-5}[0.5pt/2pt]
\noalign{\smallskip}

HD~158716 & 07-Apr-25 & 0.8977 (ND) & $0.37 \pm 0.07$ & \multirow{2}{*}{$0.13 \pm 0.02$} \\
          & 16-Jul-14 & 0.9350 (ND) & $0.16 \pm 0.03$ &  \\
\hline
\end{tabular}

\tablefoot{
Columns list the star name, the date of observation, the FAP value and detection status, the upper limit of the dipolar field strength in gauss, and the combined upper limit when multiple usable observations of the same target are retained. The parameters employed for the LSD Stokes~$I$ fits involved in the calculation are listed in Table \ref{tab:I_fit_parameters}.}

\label{Tab:NDs_characterisation}
\end{table*}

\subsection{Mean magnetic field modulus of HD~63843}

We also inspected the spectra of HD~63843 and BD~+01~1920 for evidence of resolved magnetic splitting. This check was motivated by the small central structure visible in their LSD Stokes~$V$ signatures, since such a feature can arise either from the magnetic field geometry or from magnetic splitting of spectral lines. When the magnetic field is sufficiently strong and the projected rotational velocity is low, the Zeeman effect can become large enough for individual lines to appear visibly split in the spectra, providing a direct estimate of the mean magnetic field modulus $\langle B \rangle$. We focused on standard lines used for this type of diagnostic, namely the Fe~II lines at $6149.26$ and $5018.44\,\AA$, the Nd~III line at $6145.07$\,\AA,  and the Fe~I line at $6336.82\,\AA$, although no sufficiently clean splitting was detected for the two latter in the available spectra for any of the two candidates. 

\begin{figure*}
\centering
\includegraphics[width=17cm]{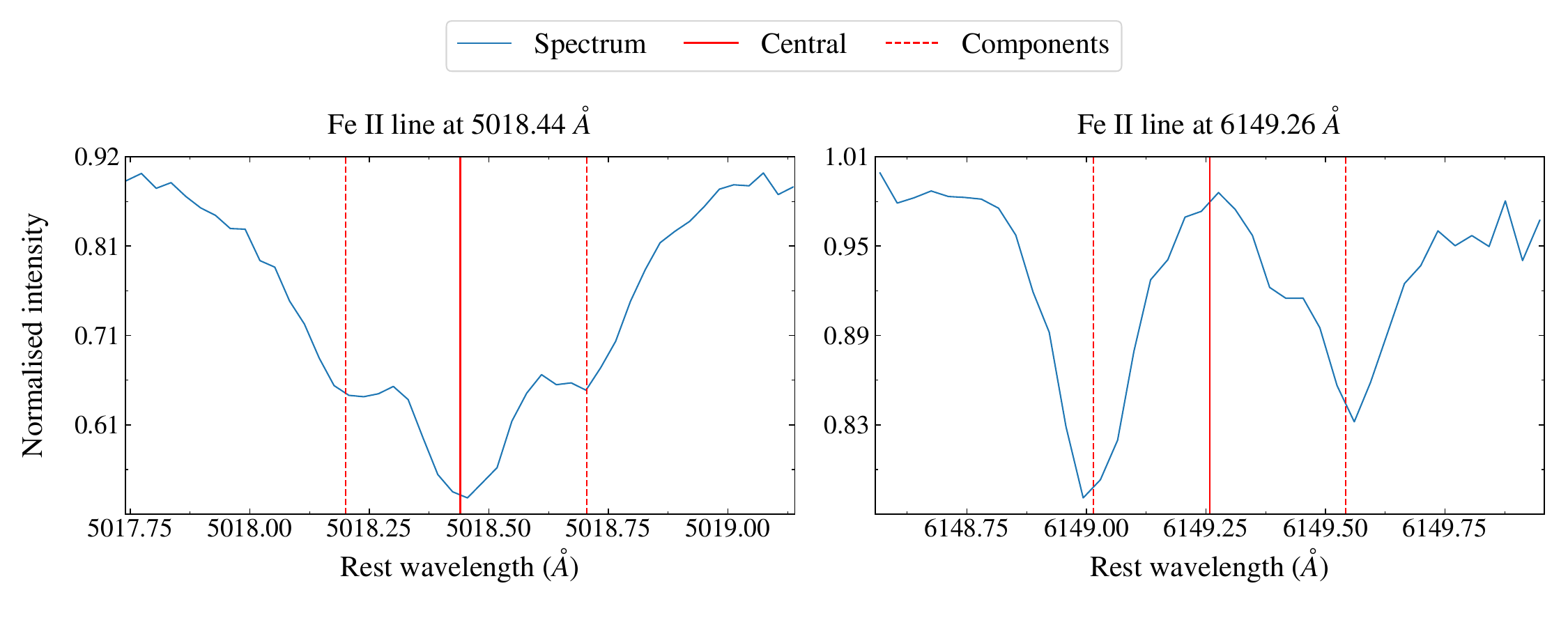}
\caption{Identification of the magnetically split Fe~II line
components in the spectrum of HD~63843. The observed normalised spectrum is
shown in blue. The vertical solid red line marks the central wavelength of
each transition, while the vertical dashed red lines indicate the identified
split components. For the $5018.44\, \AA$ line, the component
positions were determined from a three-Gaussian fit to the triplet profile.
For the $6149.26\, \AA$ line, the two component positions were
determined from centroid measurements of the blue and red absorption
features.}
\label{fig:HD63843_splitting}
\end{figure*}

For BD~+01~1920, the Fe~II lines show at most a slightly flat-bottomed profile, but no unambiguous resolved components could be identified. We therefore did not attempt to measure a mean magnetic field modulus for this star. The absence of clear splitting suggests that its surface field modulus is likely below the level at which resolved splitting is usually apparent in this diagnostic line, of order $\mathord{\sim}2$ kG \citep{Mathys17}. By contrast, HD~63843 shows clear resolved splitting in both the Fe~II lines studied, as shown in Figure~\ref{fig:HD63843_splitting}. We note that, for the $5018.44\,\AA$ line, the positions of the split components shown in the figure and listed in Table~\ref{tab:field_modulus} were obtained from a three-Gaussian fit to the triplet profile. For the $6149.26\,\AA$ line, the Gaussian fit did not reproduce the red component satisfactorily, and we therefore measured the centroids of the components instead.

\begin{table}
\caption{Magnetic splitting measurements for HD~63843 and the value of the mean magnetic field modulus obtained from them.}
\centering
\begin{tabular}{lccccc}
\hline
\noalign{\smallskip}
Line & $\lambda_b$ (\AA)& $\lambda_r$ (\AA) & $\langle B \rangle$ (kG)\\
\noalign{\smallskip}
\hline
\noalign{\smallskip}
Fe~II, 5018.44 \AA &
5018.20  &
5018.71  &
11.09 \\

Fe~II,  6149.26 \AA &
6149.01 &
6149.54 &
11.07 \\
\noalign{\smallskip}
\hline
\end{tabular}
\label{tab:field_modulus}
\end{table}

The mean magnetic field modulus is derived from the wavelength separation of the magnetic components according to $\Delta\lambda = \langle B \rangle \,{C\,\lambda_0^2}$, where $\Delta\lambda$ is the measured separation between the relevant split components, $\lambda_0$ is the wavelength of the transition, and $C$ is a line-dependent Zeeman splitting coefficient. For the $6149.26\,\AA$ line, we adopted the same values as \cite{Mathys23}, for which $C=2.70\,k$, whereas for the $5018.44\,\AA$ line, we adopted the same values as \cite{Kochukhov15}, so that $C=1.935\times 2k$. In both cases $k=4.67\times10^{-13}\,{\rm \AA^{-1}\,G^{-1}}$.

The two independent measurements give very similar values, namely $11.09$ kG from the $5018.44 \, \AA$ line and $11.07$ kG from the $6149.26\,\AA$ line. We therefore estimate a mean magnetic field modulus of $\langle B \rangle \simeq 11.1$ kG for HD~63843. This confirms that the star hosts a strong surface magnetic field and is consistent with the large longitudinal field measured from the LSD profiles.

\subsection{Upper limits for non-detections} \label{Sect: UppLims}
For the non-detections, we derived upper limits on any undetected dipolar magnetic field that could have remained hidden in the noise. To estimate these, we followed the procedure described by \cite{Neiner2015}. For each observed LSD Stokes~$V$ profile retained for the upper-limit analysis,  we generated 1000 oblique dipole models for various values of the polar magnetic field $B_{\text{pol}}$, sampling random inclination angles $i$, obliquity angles $\beta$, and rotational phases, and adding white Gaussian noise consistent with the S/N of the observations. Local Stokes~$V$ profiles were computed in the weak field approximation using the fitted LSD Stokes~$I$ profile and integrated over the visible stellar surface, using the same mean Landé factor and reference wavelength as in the corresponding observation, and a linear limb-darkening coefficient $u$ taken from \cite{Claret2019}. The parameters of the double-Gaussian least-squares fit of the LSD Stokes~$I$ profile are listed in Table \ref{tab:I_fit_parameters} and the corresponding plots are shown in Figure \ref{fig:I_fit}. 

We then applied the Neyman–Pearson likelihood ratio test to determine whether each synthetic profile would be detected, with a FAP threshold of $10^{-3}$, consistent with \cite{Donati1992}. The field strength for which 90\% of the models were detected was taken as the upper limit for that star. Following \cite{Neiner2015}, we conservatively set the uncertainty on the upper limits to $20 \%$ based on the comparison between upper limits derived from various fits of the same profile. 

We note that, for HD~65900, we excluded the lowest-S/N observation from the combined upper-limit calculation. This observation is retained in the observational log of Table \ref{Tab:observations}, and its LSD profiles in Figure \ref{fig:LSD_plots}, but its LSD Stokes~$V$ profile provides a much poorer constraint than the other two observations and was therefore not used to derive the final combined limit.

\begin{figure*}
\centering
\includegraphics[width=18cm]{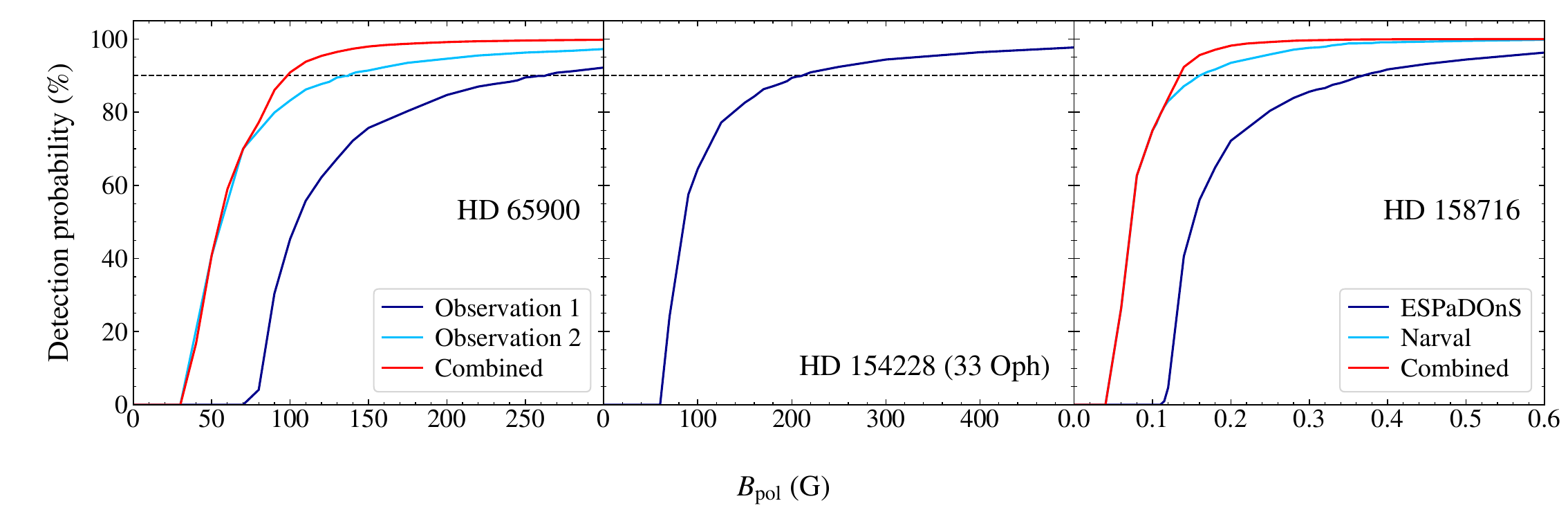}
\caption{Rate of magnetic detections among the 1000 models as a function of the dipolar field strength for all the usable observations for which no field was detected (blue lines) with the corresponding combined probability curves (red lines) when multiple observations are retained. Upper limits for the magnetic field were obtained by intersecting these curves with the 90\% probability threshold (horizontal dashed black lines).}
\label{fig: Upperlimits}
\end{figure*}

Since two of the non-detections (HD~65900 and HD~158716) had more than one usable observation, 
we combined the individual detection probabilities to obtain a stricter upper limit that accounts for the absence of a magnetic signature in the retained observations. Indeed, if we express the detection probability for observation $i$ in percent ($P_i$), the probability of detecting the field in at least one observation is 

\begin{equation}
    P_{\text{combined}}(B_{\text{pol}})= 100 \left[ 1- \prod_{i=1}^n \frac{(100-P_i(B_{\text{pol}}))}{100}\right], 
    \label{Eq: combined prob}
\end{equation}

{\noindent
where $n$ is the number of observations. To obtain a continuous combined detection probability as a function of field strength, we interpolated the discrete probability points for each observation using a cubic spline. We show in Figure \ref{fig: Upperlimits} the rate of detections among the 1000 models as a function of the field strength for each individual observation of the non-detections (blue lines), as well as the corresponding combined probability curves (red lines) in the case of multiple observations. }

Intersecting the individual probability curves in Figure \ref{fig: Upperlimits} with the 90\% probability threshold (horizontal dashed black lines), we obtained the individual upper limits $B_{\text{pol,max}}$ listed in Table \ref{Tab:NDs_characterisation}. As already mentioned in this section, we set the uncertainty of those values to $20 \%$; however, this reasoning does not apply to the uncertainties in $B_{\text{pol,max}}^{\text{comb}}$, for these combined limits were obtained using a different procedure. The combined upper limit uncertainty was obtained using the Monte Carlo method. Ten thousand independent realisations of the detection probability curves were generated by applying random horizontal shifts of the individual curves for each observation, drawn from a normal distribution with a relative uncertainty of 20\%. For each realisation, the combined detection probability was computed and the magnetic field strength at which each curve reached 90\% was determined by the corresponding intersection. The values reported in Table \ref{Tab:NDs_characterisation} correspond to the median of the distribution of intersection values, while the associated uncertainty is given by the central 68\% confidence interval derived from the corresponding percentiles.

\subsection{Critical magnetic field strength for non-detections} \label{Sect: B_crit}

Although no magnetic field was formally detected in HD~65900, HD~154228, or HD~158716, these stars exhibit several indirect features compatible with those typically associated with magnetic Ap stars: chemical peculiarities, slow rotation, and, in the case of HD~65900, variability in the TESS light curve, which is discussed further in the next section. These indicators alone, however, do not establish an Ap classification. 

As noted in Section~\ref{Sect: SlowRot}, the \citealias{Royer2014}{R14} abundances establish these objects as CP, but do not necessarily allow for an unambiguous distinction between the Am and Ap classes. The spectropolarimetric upper limits derived above, interpreted in light of the literature supporting the existence of the aforementioned magnetic desert, place strong constraints on organised fossil fields, but they do not by themselves exclude a weak Ap-type field (e.g., these stars could occupy the weak extreme tail of the field-strength distribution if the desert is not strictly empty). To assess whether an Ap interpretation remains physically plausible, we estimated the critical field strength required to suppress differential rotation in their radiative zones. If a field at the observational upper-limit level is still too weak to enforce the standard Ap-type fossil-field mechanism, then an Ap classification is disfavoured on physical grounds.

We considered two complementary approaches. The Zahn criterion \citep{Zahn2011, Mathis2005}, derived from the angular momentum transport equation, provides the mean critical field strength in the radiative zone needed to suppress differential rotation, namely

\begin{equation}
B_{\text{crit}}^{\text{Z}} = \left( 4\pi\rho \:\frac{ \, R^2 \, \Omega}{t_{\text{MS}}} \right)^{1/2}, 
\end{equation}
where $\rho$ is the mean stellar density, $R$ the stellar radius, $t_{\text{MS}}$ the main-sequence lifetime, and $\Omega$ the angular velocity.

The Spruit criterion \citep{Spruit1999}, based on fluid dynamics and magnetic diffusion, estimates the critical initial field strength above which the field remains non-axisymmetric and enforces nearly uniform rotation. This field strength is given by

\begin{equation}
B_{\text{crit, init}}^{\text{S}} = \sqrt{4 \pi \tilde \rho} \:r \left( \frac{\eta \, q^2 \, \omega^2}{3\pi^2 \, r^2} \right)^{1/3}. 
\label{eq: Spruit}
\end{equation}
Here, $\tilde \rho$ is the density in the radiative zone, $r$ a representative radius within this region, $\omega$ the angular velocity, $\eta$ the magnetic diffusivity, and $q$ the degree of differential rotation.

These calculations are intended to provide order-of-magnitude estimates of the critical fields. For the Zahn criterion, the stellar masses, radii, and main-sequence lifetimes adopted are listed in Table~\ref{tab:Bcrit}. For the Spruit criterion, we set $q=1$ and approximated the onset of the radiative zone at $0.2\,R_*$. We then took $r=0.6\,R_*$ as the midpoint of the radiative envelope for an A1-type star and used $\tilde \rho$ as the mean density of the radiative envelope for each star. The magnetic diffusivity was taken as $\eta = 3.5 \times 10^{12}\,T^{-3/2}$ cm$^2$s$^{-1}$ \citep{Spitzer1962}, which for typical A1-star temperatures gives $\eta \approx 4 \times 10^6 $ cm$^2$s$^{-1}$. These values are approximate and may differ by a modest factor from detailed MHD models, such as those by \cite{Augustson2011}. Angular velocities $\Omega$ and $\omega$ were estimated from the equatorial rotation velocity derived from the measured $v \sin{i}$. In the absence of individual inclination constraints, we assumed a random orientation of rotation axes, adopting $\sin{i} = 2/\pi \approx 0.637$ to recover equatorial velocities. Finally, to convert the internal critical fields to surface values, we applied a scaling factor for the ratio of internal to surface fields, inferred from \cite{Braithwaite2008}, which corresponds to a reduction by roughly a factor of 15.

Additionally, since expression \eqref{eq: Spruit} provides only the critical initial field, we evolved this value assuming magnetic flux conservation. The resulting quantity, $B_{\text{today}}^{\text{S}}$, does not represent the true current critical field; rather, it corresponds to the field strength today if the star had initially possessed the critical field computed from equation~\eqref{eq: Spruit}. We note that the assumption of magnetic flux conservation is a first-order approximation. In reality, some magnetic flux decay occurs \citep{Shultz2019, Landstreet08}, but this effect can be neglected for the purpose of these order-of-magnitude estimates. The calculation of such a present-day value requires the stellar radius at the initial time, which is provided in Table~\ref{tab:Bcrit}.

\begin{table*}
\captionsetup{justification=raggedright,singlelinecheck=false}
\caption{Stellar parameters and critical magnetic field strengths for the non-detections.}
\centering
\begin{tabular}{l c c c c c c c c}
\hline
\noalign{\smallskip}
Star ID 
& $M$ ($M_\odot$)
& $R$ ($R_\odot$)
& $\tau$ (Myr)
& $R_0$ ($R_\odot$)
& $\tau_0$ (Myr)
& $B_{\text{crit}}^{\text{Z}}$ (G)
& $B_{\text{crit, init}}^{\text{S}}$ (G)
& $B_{\text{today}}^{\text{S}}$ (G) \\
\noalign{\smallskip}
\hline\hline
\noalign{\smallskip}
HD 65900            & 2.42 & 2.59 & 406 & 1.72 & 24 & 1 & 73  & 32 \\
HD 154228 (33 Oph)  & 2.46 & 1.91 & 205 & 1.72 & 24 & 2 & 151 & 122 \\
HD 158716           & 2.32 & 1.80 & 462 & 1.68 & 27 & 1 & 43  & 39 \\
\noalign{\smallskip}
\hline
\end{tabular}

\tablefoot{
The table shows the adopted stellar properties (mass, radius, age, and their zero age main sequence values) with the critical magnetic field strengths derived using the Zahn ($B_{\text{crit}}^{\text{Z}}$) and Spruit criteria (initial value $B_{\text{crit, init}}^{\text{S}}$ and present-day value $B_{\text{today}}^{\text{S}}$, assuming magnetic flux conservation). Given the approximate nature of the estimated parameters used in the calculation, magnetic field values in gauss were rounded to the nearest integer.
The stellar mass and radius are taken from \cite{TESSInputs} and stellar ages from \cite{David2015}. The zero age main sequence radius ($R_0$) and age ($\tau_0$) correspond to the closest models from \cite{Mowlavi2012} assuming solar metallicity.}

\label{tab:Bcrit}
\end{table*}

The resulting surface critical field strengths are listed in Table~\ref{tab:Bcrit}. For HD~154228 and HD~65900, the upper limits comfortably exceed both critical values, preventing any meaningful constraint from this perspective. The case of HD~158716 is more informative, as its combined upper limit is considerably more stringent. For this star, a hypothetical magnetic field weak enough to have escaped detection would fall well below the Spruit critical value, and it would also lie below the Zahn critical field. This means that any magnetic field weak enough to have escaped detection in this star would be insufficient to enforce rigid rotation or suppress differential rotation in its radiative zone under either criterion. Consequently, such a weak field could not drive the standard Ap-type  mechanism for its chemical peculiarities. The implications of these results for the nature of the non-detections are discussed in the next section, where they are combined with the rest of the available information on these targets.

\section{Discussion}

In this section, we first discuss the results from the analysis of those stars with magnetic detections and then address the results for the non-detections. The two cases need different approaches.

\subsection{Magnetic detections}

Within our sample of seven low $v \sin{i}$ targets, the spectropolarimetric analysis presented in this work leads to the clear detection of magnetic fields in four stars. These detections exhibit longitudinal magnetic field strengths ranging from several hundred to several thousand gauss, as displayed in Table \ref{Tab:DDs_characterisation}. 

\begin{figure}
\centering
\includegraphics[width=8cm]{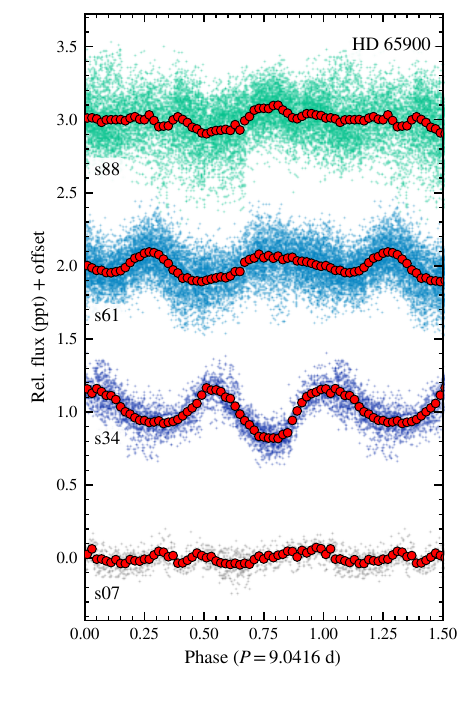}
\caption{TESS photometric light curves of HD 65900, phase-folded with respect to the rotation period. Small points show the individual normalised flux measurements from the different TESS sectors, while the red points show phase-binned mean values. The sector-to-sector changes in amplitude indicate temporal variability of the modulation, with the latest and highest-cadence sector not corresponding to the largest observed amplitude.}
\label{fig: TESS_HD65900}
\end{figure}

It is worth emphasising that all magnetic detections reported in this work originate from the target selection strategy developed by \citealias{Thomson-Paressant2024}{TP24} and detailed in Section \ref{Sect: selection of targets}.  The success of this approach in identifying all magnetic stars in our sample supports its effectiveness in the low $v \sin i$ regime. HD~65900, selected via the \citealias{Royer2014}{R14} criterion rather than this method, nonetheless displays clear rotational modulation in its TESS light curve, as illustrated in Figure \ref{fig: TESS_HD65900} and discussed further in the following subsection. This star, for which the original \citealias{Royer2014}{R14} abundance pattern did not allow an unambiguous CP subtype assignment, therefore shares one of the key indicators with the stars of the 
\citealias{Thomson-Paressant2024}{TP24} selection. The MILES \citep{MILES} medium-resolution spectrum of HD~65900, however, shows no sign of the Hümmerich depression at 5200~\AA, which is the third ingredient of this criterion. 
The non-detection of a magnetic signature in HD~65900 thus underlines the importance of requiring all three indicators of the \citealias{Thomson-Paressant2024}{TP24} criterion (adequate chemical abundance pattern, rotational modulation, and 5200~\AA\ depression) for reliable Ap selection. 
The combination, in this star, of rotational modulation, an initially ambiguous abundance pattern, the absence of the 5200~\AA\ depression, and the lack of a detected magnetic signature motivates a more detailed discussion of HD~65900 alongside the other non-detections, which is the subject of the following subsection.

In addition, it should be noted that among the four magnetic stars analysed in this work, the target HD~63843 shows evidence of pulsations: its TESS light curve displays signals consistent with those of a $\delta$~Scuti pulsator, as discussed by \citealias{Thomson-Paressant2024}{TP24}. This star therefore constitutes a promising candidate for future magneto-asteroseismic studies. Such investigations could place strong seismic constraints on the interplay between rotation and the fossil magnetic field, thereby providing valuable insight into the star’s internal structure and evolutionary state.

\subsection{Non-detections}

For the three non-detections, multiple lines of evidence converge to constrain their nature: the magnetic upper limits derived in Section \ref{Sect: UppLims}, the critical field analysis of Section \ref{Sect: B_crit}, the chemical abundance information from \citealias{Royer2014}{R14} introduced in Section \ref{Sect: SlowRot}, and, for HD 65900, additional abundance constraints and the photometric variability observed in the TESS light curve\footnote{HD~154228 was observed by TESS but shows no appreciable variability, and HD~158716 has not been observed by TESS.}. We discuss each of these strands in turn before bringing them together.

The $B_{\text{pol}}$ upper limit derived for HD 154228 ($\mathord{\sim}210$ G) from the single available observation is compatible with a relatively weak field, still within or below the typical range for Ap stars \citep{Lignieres2014}. With this single observation, it is not possible to confirm or rule out the presence of such a field; the star could host a weak Ap-type field, an ultra-weak field or no magnetic field at all. 

By contrast, the combined upper limits obtained for HD 65900 ($\mathord{\sim}98$ G) and HD 158716 ($\mathord{\sim}0.13$ G) lie below, or close to, the aforementioned usual lower boundary of the Ap magnetic regime. The constraint is especially stringent for HD~158716, whose limit reaches the sub-gauss regime. For HD~65900, the upper limit is less restrictive, but still rules out a normal large-scale Ap-like field above the magnetic desert. Although the magnetic desert discussed in the literature is an empirical regularity rather than a strict physical boundary, the position of these upper limits is already suggestive of a non-Ap nature for these stars. 

The critical-field analysis presented in Section \ref{Sect: B_crit} provides a complementary constraint, but its implications differ from star to star. For HD~158716, the upper limit on the polar field strength is below the Zahn critical field and well below the Spruit critical values. Any field weak enough to have escaped detection in this star would therefore be insufficient to enforce rigid rotation or to suppress differential rotation in its radiative zone. This disfavours the standard Ap-type mechanism as the origin of its chemical peculiarities. For HD~154228 and HD~65900, the upper limits are too high for the Zahn and Spruit thresholds to provide similarly informative constraints.  Additional observations would therefore be required to determine whether these stars host fields below or above the critical thresholds.

The abundance information provides a third, independent constraint. The \citealias{Royer2014}{R14} analysis revealed a Sc deficit in all three stars, a significant Ca deficit in HD~158716 only, and iron-peak, heavy-element, and Si overabundances in all three. The Ca and Sc deficits are characteristic of the Am class, in which atomic diffusion preferentially depletes these elements while enhancing iron-peak and heavy elements at the surface. The Si overabundance, considered in isolation, would be more suggestive of an Ap classification. The resulting abundance pattern is therefore mixed, and the absence of exhaustive rare-earth diagnostics by \citealias{Royer2014}{R14} leaves room for interpretation. 

This ambiguity is particularly important for HD~65900. Its magnetic upper limit is compatible with a sub-desert field, and its TESS light curve shows rotational modulation, so an Ap interpretation at the weak-field tail cannot be dismissed from the magnetic and photometric information alone. We therefore performed a targeted manual abundance reanalysis of Ca, Sc, and Fe for this star using \textsc{Synspec} \citep{Synspec} and updated atomic data from the VALD \citep{VALD3} and NIST \citep{NIST} databases. This analysis finds Ca to be overabundant by a factor of about 3 relative to solar from the non-resonant Ca line at 5019.97~\AA, Sc to be underabundant by a factor of about 0.1--0.8 from the lines at 4246.81 and 5526.81~\AA, and Fe to be overabundant by a factor of about 3--4. The stronger Sc line at 5031.02~\AA\ is blended with Fe~II and was therefore not used. This Ca/Sc/Fe pattern supports an Am classification for HD~65900.

HD~65900 still deserves particular attention because of its TESS variability. At the time of target selection, only the first two TESS sectors available for this star, Sectors~7 and~34, had been inspected, and the candidate rotational modulation was inferred from the Sector~34 light curve. Since then, two additional sectors, Sectors~61 and~88, have become available. Given the magnetic non-detection, we re-extracted and reanalysed the TESS photometry for all four sectors from the Full Frame Images. For each sector, we downloaded a $40\times40$-pixel region centred on the target and performed simple aperture photometry using an aperture radius of 5 pixels, with a threshold of 10 times the mean background level. The resulting light curves were detrended using 10 PCA components. We also checked alternative detrending choices, including background subtraction and fewer PCA components. The same sector-dependent trends are visible in all versions of the light curves, while the 10-component PCA correction provided the cleanest removal of instrumental systematics without suppressing the slow modulation interpreted as rotational variability.

The reanalysed light curves, shown in Figure~\ref{fig: TESS_HD65900}, confirm that the photometric variability is not stable from sector to sector. Stable, periodic photometric variability from rigid surface abundance spots tied to a strong large-scale magnetic field is the hallmark photometric signature of Ap stars, and the $\mathord{\sim}9.04$-day periodicity in HD~65900 could initially appear consistent with this picture. However, the changing amplitude and morphology of the signal argue against a stable Ap-like spot pattern. We also checked whether the variability could be caused by contamination. Within the adopted aperture, there is one contaminating star, BD~+05~1857B, with $V=12.4$, about 6.8 mag fainter than HD~65900, contributing only $\mathord{\sim}0.19\%$ of the flux of the target. The peak-to-peak amplitude in Sector~34 is about 0.4 ppt; if HD~65900 were constant, the contaminant would therefore need to vary with an intrinsic maximum-to-minimum amplitude of order 20\% to explain the observed signal. Such amplitudes are possible in some specific cases, such as eclipsing binaries or high-amplitude pulsators, but the timescale, morphology, and sector-to-sector behaviour make this interpretation unlikely. We also visually inspected the pixel-level light curves and periodograms for all sectors. They show no evidence that stars outside the adopted aperture contribute to the observed astrophysical signal. We therefore attribute the variability to HD~65900 itself.

Considering the long period, the absence of additional photometric signals, the time-variable nature of the modulation, and the lack of evidence for contamination, the behaviour is most consistent with changing rotational modulation rather than pulsation, binary-induced variability, or a processing artefact. The variability may therefore be more naturally interpreted in terms of Am-like surface inhomogeneities, potentially associated with subsurface convection or other superficial structural features sometimes invoked for Am stars \citep[e.g.][]{Blazere2018}, rather than with the stable spot configuration typical of magnetic Ap stars.

Taken together, these strands of evidence point more naturally toward an Am classification for the three non-detections, making them more plausibly Am stars than Ap stars. For HD~158716, this conclusion is supported not only by its compatible abundance pattern, but also by the magnetic constraints, since both the upper limit and the critical-field comparison are highly restrictive. For HD~65900, the absence of the 5200~\AA\ depression, the unstable rotational modulation, and the new Ca/Sc/Fe abundance reanalysis all favour an Am interpretation, despite the less constraining revised magnetic upper limit. Finally, HD~154228, for which the magnetic upper limit is the least constraining and no clear TESS variability is detected, also shows an abundance pattern compatible with an Am nature.

The contrast between the two selection strategies used in this work is itself informative. All four stars selected via the \citealias{Thomson-Paressant2024}{TP24} criterion show clear magnetic detections, while all three stars selected via the \citealias{Royer2014}{R14} criterion are best described as Am stars without detected magnetic fields. Although the sample is small, this contrast suggests that the slow rotation combined with  chemical peculiarity alone, as in the \citealias{Royer2014}{R14} criterion, is not sufficient to distinguish Am and Ap stars, whereas the addition of the 5200~\AA\ spectral depression and stable TESS rotational modulation, as in the \citealias{Thomson-Paressant2024}{TP24} criterion, appears to  discriminate more reliably in favour of Ap-type magnetism. 

These results are therefore aligned with the existing literature supporting the existence of the magnetic desert. Rather than providing clear evidence for a weak-field tail of the observed Ap-star population, the non-detections studied here appear more naturally explained as Am stars whose chemical peculiarities and, in the case of HD~65900, photometric variability arise without the strong, organised fossil fields characteristic of Ap stars. This has direct implications for target selection in upcoming surveys: criteria based only on slow rotation and chemical peculiarity may select a mixed Am/Ap population, whereas additional diagnostics tied more directly to Ap phenomenology are likely required to identify genuine magnetic Ap candidates efficiently.

\section{Conclusion}
We presented a high-resolution spectropolarimetric analysis of seven slowly rotating CP A-type stars, selected through two complementary strategies: three targets drawn from the abundance-based study of \citealias{Royer2014}{R14}, and four targets selected via the  \citealias{Thomson-Paressant2024}{TP24} criterion combining the 5200~\AA\ spectral depression with rotational modulation in TESS photometry. Magnetic fields were clearly detected in all \citealias{Thomson-Paressant2024}{TP24} targets, with longitudinal field strengths ranging from $\mathord{\sim}145$ G to $\mathord{\sim}2900$ G, while none of the three \citealias{Royer2014}{R14} targets exhibited a Zeeman signature.

For the three non-detections, we derived upper limits on the field strength, combining individual probability curves where multiple observations were available. The resulting combined upper limit for HD~158716 ($\mathord{\sim}0.13$ G) falls below the magnetic desert and below the critical field strengths derived from the Zahn and Spruit criteria, implying that any field weak enough to have escaped detection in this star would be unable to sustain the standard Ap-type fossil field mechanism. For HD~65900 and HD~154228 the upper limits ($\mathord{\sim}98$ and $210$ G respectively) are less restrictive. Together with the abundance patterns reported by \citealias{Royer2014}{R14} and, in the case of HD~65900, with the unstable rotational modulation observed in its TESS light curve, these results favour an Am classification for the three non-detections.

The contrast between the two selection strategies, although based on a small sample, is revealing: slow rotation and chemical peculiarity alone, as adopted by \citealias{Royer2014}{R14}, do not reliably discriminate between Am and Ap stars, whereas the addition of the 5200~\AA\ depression and stable TESS rotational modulation, as in the analysis by \citealias{Thomson-Paressant2024}{TP24}, appears to select Ap stars with high efficiency. This conclusion is consistent with the existing literature supporting the existence of the magnetic desert, and provides a practical guideline for target selection in upcoming surveys.

Among the four magnetic detections, HD~63843 hosts a well-characterised magnetic field (with both a longitudinal field measurement and an independent mean field modulus estimate from magnetic splitting), and exhibits $\delta$~Scuti pulsations in its TESS light curve, making it a prime candidate for magneto-asteroseismic studies, which can place strong constraints on the interplay between rotation, fossil magnetism, and stellar structure. Future spectropolarimetric follow-up of the non-detections, complemented by a dedicated abundance reanalysis including comprehensive rare-earth diagnostics, would allow for a definitive classification of HD~65900, HD~154228, and HD~158716, and would further refine our understanding of the boundary between the Ap and Am populations among slowly rotating A-type stars.


\section*{Data availability}

All ESPaDOnS spectra analysed in this work are publicly available through the CFHT Science Archive at
\url{https://www.cadc-ccda.hia-iha.nrc-cnrc.gc.ca/en/cfht/}.
The Narval observation used in this study is publicly available through the PolarBase archive at
\url{https://www.polarbase.ovgso.fr/}.

\begin{acknowledgements}
This work is based on observations obtained at the Canada-France-Hawai'i Telescope (CFHT) which is operated by the National Research Council of Canada, the Institut National des Sciences de l'Univers of the Centre National de la Recherche Scientifique of France, and the University of Hawai'i. CFHT is located on Maunakea on Hawai'i Island, a mountain of considerable cultural, natural, and ecological significance. Maunakea is a sacred site to Native Hawaiians, also known as Kānaka 'Ōiwi. Quality observations are made possible by relentless effort of the entire staff at Canada-France-Hawai'i Telescope. This work relies as well on observations obtained at the Télescope Bernard Lyot (TBL, Pic du Midi, France), which is operated by the Observatoire Midi-Pyrénées, Université de Toulouse, Centre National de la Recherche Scientifique (France). This research has made use of the SIMBAD database and the VizieR catalogue access tool, operated at CDS, Strasbourg (France), and of NASA's Astrophysics Data System (ADS). This work has made use of the VALD database, operated at Uppsala University, the Institute of Astronomy RAS in Moscow, and the University of Vienna. This paper includes data collected by the TESS mission. Funding for the TESS mission is provided by the NASA's Science Mission Directorate. J. Labadie-Bartz was funded/co-funded by the European Union through the ERC project MAGNIFY, Project No. 101126182. Views and opinions expressed are, however, those of the author(s) only and do not necessarily reflect those of the European Union or the European Research Council. Neither the European Union nor the granting authority can be held responsible for them. K. Thomson-Paressant gratefully acknowledges UK Research and Innovation (UKRI) in the form of a Frontier Research grant under the UK government's ERC Horizon Europe funding guarantee (SYMPHONY; PI Bowman; grant number: EP/Y031059/1).

\end{acknowledgements}

\bibliographystyle{aa} 
\bibliography{Astars_deFrutosRull} 

@INPROCEEDINGS{alecian2023,
       author = {{Alecian}, Georges},
        title = "{Abundance anomalies and atomic diffusion in chemically peculiar stars}",
    booktitle = {EAS2023, European Astronomical Society Annual Meeting},
         year = 2023,
        month = jul,
          eid = {401},
        pages = {401},
       adsurl = {https://ui.adsabs.harvard.edu/abs/2023eas..conf..401A}
}

@INPROCEEDINGS{schneider2024,
       author = {{Schneider}, Fabian R.~N. and {Ohlmann}, Sebastian T. and {Podsiadlowski}, Philipp and {R{\"o}pke}, Friedrich K. and {Balbus}, Steven A. and {Pakmor}, R{\"u}diger and {Springel}, Volker},
        title = "{Magnetic massive stars from stellar mergers}",
    booktitle = {Massive Stars Near and Far},
         year = 2024,
       editor = {{Mackey}, Jonathan and {Vink}, Jorick S. and {St-Louis}, Nicode},
       series = {IAU Symposium},
       volume = {361},
        month = jan,
        pages = {212-217},
          doi = {10.1017/S1743921322002794},
       adsurl = {https://ui.adsabs.harvard.edu/abs/2024IAUS..361..212S}
}

@INPROCEEDINGS{Royer2014,
       author = {{Royer}, F. and {Gebran}, M. and {Monier}, R. and {Hill}, G. and {Gulliver}, A. and {Adelman}, S. and {Smalley}, B. and {Pintado}, O. and {Reiners}, A.},
        title = "{Normal A0-A1 stars with low v sin i}",
    booktitle = {Putting A Stars into Context: Evolution, Environment, and Related Stars},
         year = 2014,
       editor = {{Mathys}, Gautier and {Griffin}, Elizabeth R. and {Kochukhov}, Oleg and {Monier}, Richard and {Wahlgren}, Glenn M.},
        month = nov,
        pages = {265-269},
       adsurl = {https://ui.adsabs.harvard.edu/abs/2014psce.conf..265R}
}

@ARTICLE{Thomson-Paressant2024,
       author = {{Thomson-Paressant}, K. and {Neiner}, C. and {Labadie-Bartz}, J.},
        title = "{Magnetism in LAMOST CP stars observed by TESS}",
      journal = {\aap},
         year = 2024,
        month = sep,
       volume = {689},
          eid = {A208},
        pages = {A208},
          doi = {10.1051/0004-6361/202450651},
archivePrefix = {arXiv},
       eprint = {2406.11554},
 primaryClass = {astro-ph.SR},
       adsurl = {https://ui.adsabs.harvard.edu/abs/2024A&A...689A.208T}
}

@ARTICLE{Hummerich2020,
       author = {{H{\"u}mmerich}, S. and {Paunzen}, E. and {Bernhard}, K.},
        title = "{A plethora of new, magnetic chemically peculiar stars from LAMOST DR4}",
      journal = {\aap},
         year = 2020,
        month = aug,
       volume = {640},
          eid = {A40},
        pages = {A40},
          doi = {10.1051/0004-6361/202037750},
archivePrefix = {arXiv},
       eprint = {2005.14444},
 primaryClass = {astro-ph.SR},
       adsurl = {https://ui.adsabs.harvard.edu/abs/2020A&A...640A..40H}
}

@INPROCEEDINGS{Donati_ESPaDOnS,
       author = {{Donati}, J.-F. and {Catala}, C. and {Landstreet}, J.~D. and {Petit}, P.},
        title = "{ESPaDOnS: The New Generation Stellar Spectro-Polarimeter. Performances and First Results}",
    booktitle = {Solar Polarization 4},
         year = 2006,
       editor = {{Casini}, R. and {Lites}, B.~W.},
       series = {Astronomical Society of the Pacific Conference Series},
       volume = {358},
        month = dec,
        pages = {362},
       adsurl = {https://ui.adsabs.harvard.edu/abs/2006ASPC..358..362D}
}

@article{Shultz2019,
   title={The magnetic early B-type stars – III. A main-sequence magnetic, rotational, and magnetospheric biography},
   volume={490},
   ISSN={1365-2966},
   url={http://dx.doi.org/10.1093/mnras/stz2551},
   DOI={10.1093/mnras/stz2551},
   number={1},
   journal={\mnras},
   publisher={Oxford University Press (OUP)},
   author={Shultz, M E and Wade, G A and Rivinius, Th and Alecian, E and Neiner, C and Petit, V and Owocki, S and ud-Doula, A and Kochukhov, O and Bohlender, D and Keszthelyi, Z},
   year={2019},
   month=sep, pages={274–295} }

@INPROCEEDINGS{Auriere_Narval,
       author = {{Auri{\`e}re}, M.},
        title = "{Stellar Polarimetry with NARVAL}",
    booktitle = {EAS Publications Series},
         year = 2003,
       editor = {{Arnaud}, J. and {Meunier}, N.},
       series = {EAS Publications Series},
       volume = {9},
        month = jan,
    publisher = {EDP},
        pages = {105},
       adsurl = {https://ui.adsabs.harvard.edu/abs/2003EAS.....9..105A}
}

@ARTICLE{Donati1997,
       author = {{Donati}, J.-F. and {Semel}, M. and {Carter}, B.~D. and {Rees}, D.~E. and {Collier Cameron}, A.},
        title = "{Spectropolarimetric observations of active stars}",
      journal = {\mnras},
         year = 1997,
        month = nov,
       volume = {291},
       number = {4},
        pages = {658-682},
          doi = {10.1093/mnras/291.4.658},
       adsurl = {https://ui.adsabs.harvard.edu/abs/1997MNRAS.291..658D}
}

@INPROCEEDINGS{Martioli2011,
       author = {{Martioli}, Eder and {Teeple}, D. and {Manset}, Nadine},
        title = "{CFHT data processing and calibration ESPaDOnS pipeline: Upena and OPERA (optical spectropolarimetry)}",
    booktitle = {Telescopes from Afar},
         year = 2011,
       editor = {{Gajadhar}, S. and {Walawender}, J. and {Genet}, R. and {Veillet}, C. and {Adamson}, A. and {Martinez}, J. and {Melnik}, J. and {Jenness}, T. and {Manset}, N.},
        month = mar,
          eid = {63},
        pages = {63},
       adsurl = {https://ui.adsabs.harvard.edu/abs/2011tfa..confE..63M}
}

@ARTICLE{Martin2018,
       author = {{Martin}, A.~J. and {Neiner}, C. and {Oksala}, M.~E. and {Wade}, G.~A. and {Keszthelyi}, Z. and {Fossati}, L. and {Marcolino}, W. and {Mathis}, S. and {Georgy}, C.},
        title = "{First results from the LIFE project: discovery of two magnetic hot evolved stars}",
      journal = {\mnras},
         year = 2018,
        month = apr,
       volume = {475},
       number = {2},
        pages = {1521-1536},
          doi = {10.1093/mnras/stx3264},
archivePrefix = {arXiv},
       eprint = {1712.07403},
 primaryClass = {astro-ph.SR},
       adsurl = {https://ui.adsabs.harvard.edu/abs/2018MNRAS.475.1521M}
}

@ARTICLE{Grunhut2017,
       author = {{Grunhut}, J.~H. and {Wade}, G.~A. and {Neiner}, C. and {Oksala}, M.~E. and {Petit}, V. and {Alecian}, E. and {Bohlender}, D.~A. and {Bouret}, J.-C. and {Henrichs}, H.~F. and {Hussain}, G.~A.~J. and {Kochukhov}, O. and {MiMeS Collaboration}},
        title = "{The MiMeS survey of Magnetism in Massive Stars: magnetic analysis of the O-type stars}",
      journal = {\mnras},
         year = 2017,
        month = feb,
       volume = {465},
       number = {2},
        pages = {2432-2470},
          doi = {10.1093/mnras/stw2743},
archivePrefix = {arXiv},
       eprint = {1610.07895},
 primaryClass = {astro-ph.SR},
       adsurl = {https://ui.adsabs.harvard.edu/abs/2017MNRAS.465.2432G}
}

@ARTICLE{Donati1992,
       author = {{Donati}, J.-F. and {Brown}, S.~F. and {Semel}, M. and {Rees}, D.~E. and {Dempsey}, R.~C. and {Matthews}, J.~M. and {Henry}, G.~W. and {Hall}, D.~S.},
        title = "{Photospheric imaging of the RS CVn system HR 1099.}",
      journal = {\aap},
         year = 1992,
        month = nov,
       volume = {265},
        pages = {682-700},
       adsurl = {https://ui.adsabs.harvard.edu/abs/1992A&A...265..682D}
}

@ARTICLE{Labadie-Bartz2023,
       author = {{Labadie-Bartz}, J. and {H{\"u}mmerich}, S. and {Bernhard}, K. and {Paunzen}, E. and {Shultz}, M.~E.},
        title = "{Photometric variability of the LAMOST sample of magnetic chemically peculiar stars as seen by TESS}",
      journal = {\aap},
         year = 2023,
        month = aug,
       volume = {676},
          eid = {A55},
        pages = {A55},
          doi = {10.1051/0004-6361/202346657},
archivePrefix = {arXiv},
       eprint = {2306.12861},
 primaryClass = {astro-ph.SR},
       adsurl = {https://ui.adsabs.harvard.edu/abs/2023A&A...676A..55L}
}

@MISC{Claret2019,
       author = {{Claret}, A.},
        title = "{VizieR Online Data Catalog: Limb-darkening for Space Mission GAIA (Claret, 2019)}",
 howpublished = {VizieR On-line Data Catalog: VI/154.  Originally published in: 2019RNASS...3...17C},
         year = 2019,
        month = jan,
          eid = {VI/154},
       adsurl = {https://ui.adsabs.harvard.edu/abs/2019yCat.6154....0C}
}

@ARTICLE{Rees1979,
       author = {{Rees}, D.~E. and {Semel}, M.~D.},
        title = "{Line formation in an unresolved magnetic element: a test of the centre of gravity method.}",
      journal = {\aap},
         year = 1979,
        month = apr,
       volume = {74},
       number = {1},
        pages = {1-5},
       adsurl = {https://ui.adsabs.harvard.edu/abs/1979A&A....74....1R}
}

@ARTICLE{Wade2000,
       author = {{Wade}, G.~A. and {Donati}, J.-F. and {Landstreet}, J.~D. and {Shorlin}, S.~L.~S.},
        title = "{High-precision magnetic field measurements of Ap and Bp stars}",
      journal = {\mnras},
         year = 2000,
        month = apr,
       volume = {313},
       number = {4},
        pages = {851-867},
          doi = {10.1046/j.1365-8711.2000.03271.x},
       adsurl = {https://ui.adsabs.harvard.edu/abs/2000MNRAS.313..851W}
}

@ARTICLE{Neiner2015,
       author = {{Neiner}, C. and {Grunhut}, J. and {Leroy}, B. and {De Becker}, M. and {Rauw}, G.},
        title = "{Search for magnetic fields in particle-accelerating colliding-wind binaries}",
      journal = {\aap},
         year = 2015,
        month = mar,
       volume = {575},
          eid = {A66},
        pages = {A66},
          doi = {10.1051/0004-6361/201425193},
archivePrefix = {arXiv},
       eprint = {1412.5327},
 primaryClass = {astro-ph.SR},
       adsurl = {https://ui.adsabs.harvard.edu/abs/2015A&A...575A..66N}
}

@INPROCEEDINGS{NeinerMathis2015,
       author = {{Neiner}, Coralie and {Mathis}, St{\'e}phane and {Alecian}, Evelyne and {Emeriau}, Constance and {Grunhut}, Jason and {BinaMIcS} and {MiMeS Collaborations}},
        title = "{The origin of magnetic fields in hot stars}",
    booktitle = {Polarimetry},
         year = 2015,
       editor = {{Nagendra}, K.~N. and {Bagnulo}, Stefano and {Centeno}, Rebecca and {Jes{\'u}s Mart{\'\i}nez Gonz{\'a}lez}, Mar{\'\i}a.},
       series = {IAU Symposium},
       volume = {305},
        month = oct,
        pages = {61-66},
          doi = {10.1017/S1743921315004524},
archivePrefix = {arXiv},
       eprint = {1502.00226},
 primaryClass = {astro-ph.SR},
       adsurl = {https://ui.adsabs.harvard.edu/abs/2015IAUS..305...61N}
}

@ARTICLE{Michaud1970,
       author = {{Michaud}, Georges},
        title = "{Diffusion Processes in Peculiar a Stars}",
      journal = {\apj},
         year = 1970,
        month = may,
       volume = {160},
        pages = {641},
          doi = {10.1086/150459},
       adsurl = {https://ui.adsabs.harvard.edu/abs/1970ApJ...160..641M}
}

@ARTICLE{Fuller2019,
       author = {{Fuller}, Jim and {Piro}, Anthony L. and {Jermyn}, Adam S.},
        title = "{Slowing the spins of stellar cores}",
      journal = {\mnras},
         year = 2019,
        month = may,
       volume = {485},
       number = {3},
        pages = {3661-3680},
          doi = {10.1093/mnras/stz514},
archivePrefix = {arXiv},
       eprint = {1902.08227},
 primaryClass = {astro-ph.SR},
       adsurl = {https://ui.adsabs.harvard.edu/abs/2019MNRAS.485.3661F}
}

@ARTICLE{Briquet2012,
       author = {{Briquet}, M. and {Neiner}, C. and {Aerts}, C. and {Morel}, T. and {Mathis}, S. and {Reese}, D.~R. and {Lehmann}, H. and {Costero}, R. and {Echevarria}, J. and {Handler}, G. and {Kambe}, E. and {Hirata}, R. and {Masuda}, S. and {Wright}, D. and {Yang}, S. and {Pintado}, O. and {Mkrtichian}, D. and {Lee}, B.~C. and {Han}, I. and {Bruch}, A. and {De Cat}, P. and {Uytterhoeven}, K. and {Lefever}, K. and {Vanautgaerden}, J. and {de Batz}, B. and {Fr{\'e}mat}, Y. and {Henrichs}, H. and {Geers}, V.~C. and {Martayan}, C. and {Hubert}, A.~M. and {Thizy}, O. and {Tijani}, A.},
        title = "{Multisite spectroscopic seismic study of the {\ensuremath{\beta}} Cep star V2052 Ophiuchi: inhibition of mixing by its magnetic field}",
      journal = {\mnras},
         year = 2012,
        month = nov,
       volume = {427},
       number = {1},
        pages = {483-493},
          doi = {10.1111/j.1365-2966.2012.21933.x},
archivePrefix = {arXiv},
       eprint = {1208.4250},
 primaryClass = {astro-ph.SR},
       adsurl = {https://ui.adsabs.harvard.edu/abs/2012MNRAS.427..483B}
}

@ARTICLE{Preston1974,
       author = {{Preston}, G.~W.},
        title = "{The chemically peculiar stars of the upper main sequence.}",
      journal = {\araa},
         year = 1974,
        month = jan,
       volume = {12},
        pages = {257-277},
          doi = {10.1146/annurev.aa.12.090174.001353},
       adsurl = {https://ui.adsabs.harvard.edu/abs/1974ARA&A..12..257P}
}

@ARTICLE{Abt1995,
       author = {{Abt}, Helmut A. and {Morrell}, Nidia I.},
        title = "{The Relation between Rotational Velocities and Spectral Peculiarities among A-Type Stars}",
      journal = {\apjs},
         year = 1995,
        month = jul,
       volume = {99},
        pages = {135},
          doi = {10.1086/192182},
       adsurl = {https://ui.adsabs.harvard.edu/abs/1995ApJS...99..135A}
}

@ARTICLE{Stepien1998,
       author = {{Stepien}, K.},
        title = "{Why are magnetic AP stars slowly rotating?}",
      journal = {Contrib. Astron. Obs. Skaln. Pleso},
         year = 1998,
        month = apr,
       volume = {27},
       number = {3},
        pages = {205-212},
          doi = {10.48550/arXiv.astro-ph/9805032},
archivePrefix = {arXiv},
       eprint = {astro-ph/9805032},
 primaryClass = {astro-ph},
       adsurl = {https://ui.adsabs.harvard.edu/abs/1998CoSka..27..205S}
}

@ARTICLE{Kochukhov2006,
       author = {{Kochukhov}, O. and {Bagnulo}, S.},
        title = "{Evolutionary state of magnetic chemically peculiar stars}",
      journal = {\aap},
         year = 2006,
        month = may,
       volume = {450},
       number = {2},
        pages = {763-775},
          doi = {10.1051/0004-6361:20054596},
archivePrefix = {arXiv},
       eprint = {astro-ph/0601461},
 primaryClass = {astro-ph},
       adsurl = {https://ui.adsabs.harvard.edu/abs/2006A&A...450..763K}
}

@ARTICLE{PLATO,
       author = {{Rauer}, Heike and {Aerts}, Conny and {Cabrera}, Juan and {Deleuil}, Magali and {Erikson}, Anders and {Gizon}, Laurent and {Goupil}, Mariejo and {Heras}, Ana and {Walloschek}, Thomas and {Lorenzo-Alvarez}, Jose and {Marliani}, Filippo and {Martin-Garcia}, C{\'e}sar and {Mas-Hesse}, J. Miguel and {O'Rourke}, Laurence and {Osborn}, Hugh and {Pagano}, Isabella and {Piotto}, Giampaolo and {Pollacco}, Don and {Ragazzoni}, Roberto and {Ramsay}, Gavin and {Udry}, St{\'e}phane and {Appourchaux}, Thierry and {Benz}, Willy and {Brandeker}, Alexis and {G{\"u}del}, Manuel and {Janot-Pacheco}, Eduardo and {Kabath}, Petr and {Kjeldsen}, Hans and {Min}, Michiel and {Santos}, Nuno and {Smith}, Alan and {Suarez}, Juan-Carlos and {Werner}, Stephanie C. and {Aboudan}, Alessio and {Abreu}, Manuel and {Acu{\~n}a}, Lorena and {Adams}, Moritz and {Adibekyan}, Vardan and {Affer}, Laura and {Agneray}, Fran{\c{c}}ois and {Agnor}, Craig and {Aguirre B{\o}rsen-Koch}, Victor and {Ahmed}, Saad and {Aigrain}, Suzanne and {Al-Bahlawan}, Ashraf and {Alcacera Gil}, Ma de los Angeles and {Alei}, Eleonora and {Alencar}, Silvia and {Alexander}, Richard and {Alfonso-Garz{\'o}n}, Julia and {Alibert}, Yann and {Allende Prieto}, Carlos and {Almeida}, Leonardo and {Alonso Sobrino}, Roi and {Altavilla}, Giuseppe and {Althaus}, Christian and {Alvarez Trujillo}, Luis Alonso and {Amarsi}, Anish and {Ammler-von Eiff}, Matthias and {Am{\^o}res}, Eduardo and {Andrade}, Laerte and {Antoniadis-Karnavas}, Alexandros and {Ant{\'o}nio}, Carlos and {Aparicio del Moral}, Beatriz and {Appolloni}, Matteo and {Arena}, Claudio and {Armstrong}, David and {Aroca Aliaga}, Jose and {Asplund}, Martin and {Audenaert}, Jeroen and {Auricchio}, Natalia and {Avelino}, Pedro and {Baeke}, Ann and {Bailli{\'e}}, Kevin and {Balado}, Ana and {Ballber Balaguer{\'o}}, Pau and {Balestra}, Andrea and {Ball}, Warrick and {Ballans}, Herve and {Ballot}, Jerome and {Barban}, Caroline and {Barbary}, Ga{\"e}le and {Barbieri}, Mauro and {Barcel{\'o} Forteza}, Sebasti{\`a} and {Barker}, Adrian and {Barklem}, Paul and {Barnes}, Sydney and {Barrado Navascues}, David and {Barragan}, Oscar and {Baruteau}, Cl{\'e}ment and {Basu}, Sarbani and {Baudin}, Frederic and {Baumeister}, Philipp and {Bayliss}, Daniel and {Bazot}, Michael and {Beck}, Paul G. and {Belkacem}, Kevin and {Bellinger}, Earl and {Benatti}, Serena and {Benomar}, Othman and {B{\'e}rard}, Diane and {Bergemann}, Maria and {Bergomi}, Maria and {Bernardo}, Pierre and {Biazzo}, Katia and {Bignamini}, Andrea and {Bigot}, Lionel and {Billot}, Nicolas and {Binet}, Martin and {Biondi}, David and {Biondi}, Federico and {Birch}, Aaron C. and {Bitsch}, Bertram and {Bluhm Ceballos}, Paz Victoria and {B{\'o}di}, Attila and {Bogn{\'a}r}, Zs{\'o}fia and {Boisse}, Isabelle and {Bolmont}, Emeline and {Bonanno}, Alfio and {Bonavita}, Mariangela and {Bonfanti}, Andrea and {Bonfils}, Xavier and {Bonito}, Rosaria and {Bonomo}, Aldo Stefano and {B{\"o}rner}, Anko and {Boro Saikia}, Sudeshna and {Borreguero Mart{\'\i}n}, Elisa and {Borsa}, Francesco and {Borsato}, Luca and {Bossini}, Diego and {Bouchy}, Francois and {Bou{\'e}}, Gwena{\"e}l and {Boufleur}, Rodrigo and {Boumier}, Patrick and {Bourrier}, Vincent and {Bowman}, Dominic M. and {Bozzo}, Enrico and {Bradley}, Louisa and {Bray}, John and {Bressan}, Alessandro and {Breton}, Sylvain and {Brienza}, Daniele and {Brito}, Ana and {Brogi}, Matteo and {Brown}, Beverly and {Brown}, David J.~A. and {Brun}, Allan Sacha and {Bruno}, Giovanni and {Bruns}, Michael and {Buchhave}, Lars A. and {Bugnet}, Lisa and {Buldgen}, Ga{\"e}l and {Burgess}, Patrick and {Busatta}, Andrea and {Busso}, Giorgia and {Buzasi}, Derek and {Caballero}, Jos{\'e} A. and {Cabral}, Alexandre and {Cabrero Gomez}, Juan-Francisco and {Calderone}, Flavia and {Cameron}, Robert and {Cameron}, Andrew and {Campante}, Tiago and {Campos Gestal}, N{\'e}stor and {Canto Martins}, Bruno Leonardo and {Cara}, Christophe and {Carone}, Ludmila and {Carrasco}, Josep Manel and {Casagrande}, Luca and {Casewell}, Sarah L. and {Cassisi}, Santi and {Castellani}, Marco and {Castro}, Matthieu and {Catala}, Claude and {Catal{\'a}n Fern{\'a}ndez}, Irene and {Catelan}, M{\'a}rcio and {Cegla}, Heather and {Cerruti}, Chiara and {Cessa}, Virginie and {Chadid}, Merieme and {Chaplin}, William and {Charpinet}, Stephane and {Chiappini}, Cristina and {Chiarucci}, Simone and {Chiavassa}, Andrea and {Chinellato}, Simonetta and {Chirulli}, Giovanni and {Christensen-Dalsgaard}, J{\o}rgen and {Church}, Ross and {Claret}, Antonio and {Clarke}, Cathie and {Claudi}, Riccardo and {Clermont}, Lionel and {Coelho}, Hugo and {Coelho}, Joao and {Cogato}, Fabrizio and {Colom{\'e}}, Josep and {Condamin}, Mathieu and {Conde Garc{\'\i}a}, Fernando and {Conseil}, Simon},
        title = "{The PLATO mission}",
      journal = {Exp. Astron.},
         year = 2025,
        month = jun,
       volume = {59},
       number = {3},
          eid = {26},
        pages = {26},
          doi = {10.1007/s10686-025-09985-9},
archivePrefix = {arXiv},
       eprint = {2406.05447},
 primaryClass = {astro-ph.IM},
       adsurl = {https://ui.adsabs.harvard.edu/abs/2025ExA....59...26R}
}

@INPROCEEDINGS{Neiner2021,
       author = {{Neiner}, C. and {Labadie-Bartz}, J. and {Catala}, C. and {Bernhard}, K. and {Bowman}, D.~M. and {David-Uraz}, A. and {H{\"u}mmerich}, S. and {Paunzen}, E. and {Shultz}, M.~E.},
        title = "{MOBSTER: Magneto-asteroseismology of hot stars with TESS}",
    booktitle = {SF2A-2021: Proceedings of the Annual meeting of the French Society of Astronomy and Astrophysics},
         year = 2021,
       editor = {{Siebert}, A. and {Bailli{\'e}}, K. and {Lagadec}, E. and {Lagarde}, N. and {Malzac}, J. and {Marquette}, J. -B. and {N'Diaye}, M. and {Richard}, J. and {Venot}, O.},
        month = dec,
        pages = {161-164},
       adsurl = {https://ui.adsabs.harvard.edu/abs/2021sf2a.conf..161N}
}

@BOOK{Spitzer1962,
       author = {{Spitzer}, L.},
        title = "{Physics of Fully Ionized Gases}",
         year = 1962,
         publisher = {Interscience Publishers},
       adsurl = {https://ui.adsabs.harvard.edu/abs/1962pfig.book.....S}
}

@ARTICLE{TESSInputs,
       author = {{Stassun}, Keivan G. and {Oelkers}, Ryan J. and {Paegert}, Martin and {Torres}, Guillermo and {Pepper}, Joshua and {De Lee}, Nathan and {Collins}, Kevin and {Latham}, David W. and {Muirhead}, Philip S. and {Chittidi}, Jay and {Rojas-Ayala}, B{\'a}rbara and {Fleming}, Scott W. and {Rose}, Mark E. and {Tenenbaum}, Peter and {Ting}, Eric B. and {Kane}, Stephen R. and {Barclay}, Thomas and {Bean}, Jacob L. and {Brassuer}, C.~E. and {Charbonneau}, David and {Ge}, Jian and {Lissauer}, Jack J. and {Mann}, Andrew W. and {McLean}, Brian and {Mullally}, Susan and {Narita}, Norio and {Plavchan}, Peter and {Ricker}, George R. and {Sasselov}, Dimitar and {Seager}, S. and {Sharma}, Sanjib and {Shiao}, Bernie and {Sozzetti}, Alessandro and {Stello}, Dennis and {Vanderspek}, Roland and {Wallace}, Geoff and {Winn}, Joshua N.},
        title = "{The Revised TESS Input Catalog and Candidate Target List}",
      journal = {\aj},
         year = 2019,
        month = oct,
       volume = {158},
       number = {4},
          eid = {138},
        pages = {138},
          doi = {10.3847/1538-3881/ab3467},
archivePrefix = {arXiv},
       eprint = {1905.10694},
 primaryClass = {astro-ph.SR},
       adsurl = {https://ui.adsabs.harvard.edu/abs/2019AJ....158..138S}
}

@article{David2015,
doi = {10.1088/0004-637X/804/2/146},
url = {https://doi.org/10.1088/0004-637X/804/2/146},
year = {2015},
month = {may},
publisher = {\apj},
volume = {804},
number = {2},
pages = {146},
author = {David, Trevor J. and Hillenbrand, Lynne A.},
title = {THE AGES OF EARLY-TYPE STARS: STRÖMGREN PHOTOMETRIC METHODS CALIBRATED, VALIDATED, TESTED, AND APPLIED TO HOSTS AND PROSPECTIVE HOSTS OF DIRECTLY IMAGED EXOPLANETS},
journal = {The Astrophysical Journal}
}

@ARTICLE{Mathis2005,
       author = {{Mathis}, S. and {Zahn}, J.-P.},
        title = "{Transport and mixing in the radiation zones of rotating stars. II. Axisymmetric magnetic field}",
      journal = {\aap},
         year = 2005,
        month = sep,
       volume = {440},
       number = {2},
        pages = {653-666},
          doi = {10.1051/0004-6361:20052640},
archivePrefix = {arXiv},
       eprint = {astro-ph/0506105},
 primaryClass = {astro-ph},
       adsurl = {https://ui.adsabs.harvard.edu/abs/2005A&A...440..653M}
}

@INPROCEEDINGS{Zahn2011,
       author = {{Zahn}, Jean-Paul},
        title = "{Rapid rotation and mixing in active OB stars - Physical processes}",
    booktitle = {Active OB Stars: Structure, Evolution, Mass Loss, and Critical Limits},
         year = 2011,
       editor = {{Neiner}, Coralie and {Wade}, Gregg and {Meynet}, Georges and {Peters}, Geraldine},
       series = {IAU Symposium},
       volume = {272},
        month = jul,
        pages = {14-25},
          doi = {10.1017/S1743921311009926},
       adsurl = {https://ui.adsabs.harvard.edu/abs/2011IAUS..272...14Z}
}

@ARTICLE{Spruit1999,
       author = {{Spruit}, H.~C.},
        title = "{Differential rotation and magnetic fields in stellar interiors}",
      journal = {\aap},
         year = 1999,
        month = sep,
       volume = {349},
        pages = {189-202},
          doi = {10.48550/arXiv.astro-ph/9907138},
archivePrefix = {arXiv},
       eprint = {astro-ph/9907138},
 primaryClass = {astro-ph},
       adsurl = {https://ui.adsabs.harvard.edu/abs/1999A&A...349..189S}
}

@ARTICLE{Braithwaite2008,
       author = {{Braithwaite}, Jonathan},
        title = "{On non-axisymmetric magnetic equilibria in stars}",
      journal = {\mnras},
         year = 2008,
        month = jun,
       volume = {386},
       number = {4},
        pages = {1947-1958},
          doi = {10.1111/j.1365-2966.2008.13218.x},
archivePrefix = {arXiv},
       eprint = {0803.1661},
 primaryClass = {astro-ph},
       adsurl = {https://ui.adsabs.harvard.edu/abs/2008MNRAS.386.1947B}
}

@INPROCEEDINGS{Lignieres2014,
       author = {{Ligni{\`e}res}, Fran{\c{c}}ois and {Petit}, Pascal and {Auri{\`e}re}, Michel and {Wade}, Gregg A. and {B{\"o}hm}, Torsten},
        title = "{The dichotomy between strong and ultra-weak magnetic fields among intermediate-mass stars}",
    booktitle = {Magnetic Fields throughout Stellar Evolution},
         year = 2014,
       editor = {{Petit}, Pascal and {Jardine}, Moira and {Spruit}, Hendrik C.},
       series = {IAU Symposium},
       volume = {302},
        month = aug,
        pages = {338-347},
          doi = {10.1017/S1743921314002440},
archivePrefix = {arXiv},
       eprint = {1402.5362},
 primaryClass = {astro-ph.SR},
       adsurl = {https://ui.adsabs.harvard.edu/abs/2014IAUS..302..338L}
}

@ARTICLE{Blazere2018,
       author = {{Blaz{\`e}re}, A. and {Petit}, P. and {Neiner}, C.},
        title = "{The magnetic properties of Am stars}",
      journal = {Contrib. Astron. Obs. Skaln. Pleso},
         year = 2018,
        month = jan,
       volume = {48},
       number = {1},
        pages = {48-52},
       adsurl = {https://ui.adsabs.harvard.edu/abs/2018CoSka..48...48B}
}

@INPROCEEDINGS{Augustson2011,
       author = {{Augustson}, Kyle C. and {Brun}, Allan S. and {Toomre}, Juri},
        title = "{Convection and dynamo action in B stars}",
    booktitle = {Astrophysical Dynamics: From Stars to Galaxies},
         year = 2011,
       editor = {{Brummell}, Nicholas H. and {Brun}, A. Sacha and {Miesch}, Mark S. and {Ponty}, Yannick},
       series = {IAU Symposium},
       volume = {271},
        month = aug,
        pages = {361-362},
          doi = {10.1017/S1743921311017790},
archivePrefix = {arXiv},
       eprint = {1011.1016},
 primaryClass = {astro-ph.SR},
       adsurl = {https://ui.adsabs.harvard.edu/abs/2011IAUS..271..361A}
}

@ARTICLE{Mowlavi2012,
       author = {{Mowlavi}, N. and {Eggenberger}, P. and {Meynet}, G. and {Ekstr{\"o}m}, S. and {Georgy}, C. and {Maeder}, A. and {Charbonnel}, C. and {Eyer}, L.},
        title = "{Stellar mass and age determinations . I. Grids of stellar models from Z = 0.006 to 0.04 and M = 0.5 to 3.5 M$_{☉}$}",
      journal = {\aap},
         year = 2012,
        month = may,
       volume = {541},
          eid = {A41},
        pages = {A41},
          doi = {10.1051/0004-6361/201117749},
archivePrefix = {arXiv},
       eprint = {1201.3628},
 primaryClass = {astro-ph.SR},
       adsurl = {https://ui.adsabs.harvard.edu/abs/2012A&A...541A..41M}
}

@ARTICLE{Balona2015,
       author = {{Balona}, L.~A. and {Catanzaro}, G. and {Abedigamba}, O.~P. and {Ripepi}, V. and {Smalley}, B.},
        title = "{Spots on Am stars}",
      journal = {\mnras},
         year = 2015,
        month = apr,
       volume = {448},
       number = {2},
        pages = {1378-1388},
          doi = {10.1093/mnras/stv076},
archivePrefix = {arXiv},
       eprint = {1501.06746},
 primaryClass = {astro-ph.SR},
       adsurl = {https://ui.adsabs.harvard.edu/abs/2015MNRAS.448.1378B}
}

@ARTICLE{Auriere2007,
       author = {{Auri{\`e}re}, M. and {Wade}, G.~A. and {Silvester}, J. and {Ligni{\`e}res}, F. and {Bagnulo}, S. and {Bale}, K. and {Dintrans}, B. and {Donati}, J.~F. and {Folsom}, C.~P. and {Gruberbauer}, M. and {Hui Bon Hoa}, A. and {Jeffers}, S. and {Johnson}, N. and {Landstreet}, J.~D. and {L{\`e}bre}, A. and {Lueftinger}, T. and {Marsden}, S. and {Mouillet}, D. and {Naseri}, S. and {Paletou}, F. and {Petit}, P. and {Power}, J. and {Rincon}, F. and {Strasser}, S. and {Toqu{\'e}}, N.},
        title = "{Weak magnetic fields in Ap/Bp stars. Evidence for a dipole field lower limit and a tentative interpretation of the magnetic dichotomy}",
      journal = {\aap},
         year = 2007,
        month = dec,
       volume = {475},
       number = {3},
        pages = {1053-1065},
          doi = {10.1051/0004-6361:20078189},
archivePrefix = {arXiv},
       eprint = {0710.1554},
 primaryClass = {astro-ph},
       adsurl = {https://ui.adsabs.harvard.edu/abs/2007A&A...475.1053A}
}

@ARTICLE{MILES,
       author = {{S{\'a}nchez-Bl{\'a}zquez}, P. and {Peletier}, R.~F. and {Jim{\'e}nez-Vicente}, J. and {Cardiel}, N. and {Cenarro}, A.~J. and {Falc{\'o}n-Barroso}, J. and {Gorgas}, J. and {Selam}, S. and {Vazdekis}, A.},
        title = "{Medium-resolution Isaac Newton Telescope library of empirical spectra}",
      journal = {\mnras},
         year = 2006,
        month = sep,
       volume = {371},
       number = {2},
        pages = {703-718},
          doi = {10.1111/j.1365-2966.2006.10699.x},
archivePrefix = {arXiv},
       eprint = {astro-ph/0607009},
 primaryClass = {astro-ph},
       adsurl = {https://ui.adsabs.harvard.edu/abs/2006MNRAS.371..703S}
}

@MISC{Synspec,
       author = {{Hubeny}, Ivan and {Lanz}, Thierry},
        title = "{Synspec: General Spectrum Synthesis Program}",
 howpublished = {Astrophysics Source Code Library, record ascl:1109.022},
         year = 2011,
        month = sep,
          eid = {ascl:1109.022},
archivePrefix = {ascl},
       eprint = {1109.022},
       adsurl = {https://ui.adsabs.harvard.edu/abs/2011ascl.soft09022H}
}

@MISC{NIST,
  author = {{Kramida}, A. and {Ralchenko}, Yu. and {Reader}, J. and {NIST ASD Team}},
  title = {{NIST Atomic Spectra Database (version 5.12)}},
  year = 2024,
  howpublished = {National Institute of Standards and Technology, Gaithersburg, MD},
  note = {Available: \url{https://physics.nist.gov/asd}},
  doi = {10.18434/T4W30F}
}

@ARTICLE{VALD3,
       author = {{Ryabchikova}, T. and {Piskunov}, N. and {Kurucz}, R.~L. and 
                 {Stempels}, H.~C. and {Heiter}, U. and {Pakhomov}, Yu. and 
                 {Barklem}, P.~S.},
        title = "{A major upgrade of the VALD database}",
      journal = {\physscr},
         year = 2015,
       volume = {90},
       number = {5},
        pages = {054005},
          doi = {10.1088/0031-8949/90/5/054005},
       adsurl = {https://ui.adsabs.harvard.edu/abs/2015PhyS...90e4005R}
}

@ARTICLE{Landstreet08,
       author = {{Landstreet}, J.~D. and {Bagnulo}, S. and {Andretta}, V. and {Fossati}, L. and {Mason}, E. and {Silaj}, J. and {Wade}, G.~A.},
        title = "{Evolution of global magnetic fields in main sequence A and B stars}",
      journal = {Contrib. Astron. Obs. Skaln. Pleso},
         year = 2008,
        month = apr,
       volume = {38},
       number = {2},
        pages = {391-396},
       adsurl = {https://ui.adsabs.harvard.edu/abs/2008CoSka..38..391L}
}

@ARTICLE{Mathys17,
       author = {{Mathys}, G.},
        title = "{Ap stars with resolved magnetically split lines: Magnetic field determinations from Stokes I and V spectra{\ensuremath{\star}}}",
      journal = {\aap},
         year = 2017,
        month = may,
       volume = {601},
          eid = {A14},
        pages = {A14},
          doi = {10.1051/0004-6361/201628429},
archivePrefix = {arXiv},
       eprint = {1612.03632},
 primaryClass = {astro-ph.SR},
       adsurl = {https://ui.adsabs.harvard.edu/abs/2017A&A...601A..14M}
}

@ARTICLE{Mathys23,
       author = {{Mathys}, G. and {Khalack}, V. and {Kobzar}, O. and {LeBlanc}, F. and {North}, P.~L.},
        title = "{HD 213258: A new rapidly oscillating, super slowly rotating, strongly magnetic Ap star in a spectroscopic binary}",
      journal = {\aap},
         year = 2023,
        month = feb,
       volume = {670},
          eid = {A72},
        pages = {A72},
          doi = {10.1051/0004-6361/202245568},
archivePrefix = {arXiv},
       eprint = {2212.12752},
 primaryClass = {astro-ph.SR},
       adsurl = {https://ui.adsabs.harvard.edu/abs/2023A&A...670A..72M}
}

@ARTICLE{Kochukhov15,
       author = {{Kochukhov}, O. and {Rusomarov}, N. and {Valenti}, J.~A. and {Stempels}, H.~C. and {Snik}, F. and {Rodenhuis}, M. and {Piskunov}, N. and {Makaganiuk}, V. and {Keller}, C.~U. and {Johns-Krull}, C.~M.},
        title = "{Magnetic field topology and chemical spot distributions in the extreme Ap star HD 75049}",
      journal = {\aap},
         year = 2015,
        month = feb,
       volume = {574},
          eid = {A79},
        pages = {A79},
          doi = {10.1051/0004-6361/201425065},
archivePrefix = {arXiv},
       eprint = {1411.7518},
 primaryClass = {astro-ph.SR},
       adsurl = {https://ui.adsabs.harvard.edu/abs/2015A&A...574A..79K}
}

\newpage 

\begin{appendix}
\onecolumn

\section{Observation log}

\begin{table*}[h]
\captionsetup{justification=raggedright,singlelinecheck=false}
\caption{Observing log and signal-to-noise ratios.}
\label{Tab:observations}
\centering
\begin{tabular}{llllllll}
\hline
\noalign{\smallskip}
Target & Date & Instrument & Mid-HJD & $T_{\text{exp}}$ & S/N & S/N & S/N \\
& & & (+2455000) & (seq $\times$ 4 $\times$ s) & (Stokes~$I$, 503 nm) & (LSD~$I$) & (LSD~$V$) \\
\noalign{\smallskip}
\hline\hline
\noalign{\smallskip}
\multirow[t]{3}{*}{HD~65900}
& 04-Jan-24 & ESPaDOnS & 5314.0829 & $1\times4\times165$ & 458 & 4523 & 10727 \\
& 04-Jan-24 & ESPaDOnS & 5314.0927 & $1\times4\times165$ & 101 & 1972 & 1702 \\
& 08-Jan-24 & ESPaDOnS & 5318.0527 & $1\times4\times165$ & 802 & 5086 & 20395 \\
\noalign{\smallskip}
HD~154228
& 07-Apr-25 & ESPaDOnS & 5772.9805 & $1\times4\times377$ & 799 & 6871 & 15184 \\
\noalign{\smallskip}
\multirow[t]{2}{*}{HD~158716}
& 07-Apr-25    & ESPaDOnS & 5772.9982 & $1\times4\times300$ & 578 & 2360 & 12185 \\
& 16-Jul-14  & Narval   & 1855.4461 & $1\times4\times600$ & 420 & 2523 & 12586 \\
\noalign{\smallskip}
\cdashline{1-8}
\noalign{\smallskip}
HD~63843
& 19-Jan-24 & ESPaDOnS & 5329.0059 & $1\times4\times280$ & 115 & 1568 & 2831 \\
\noalign{\smallskip}
HD~266267
& 16-Jan-24 & ESPaDOnS & 5326.0506 & $1\times4\times55$ & 53 & 879 & 1616 \\
\noalign{\smallskip}
BD~+01~1920
& 08-Jan-24 & ESPaDOnS & 5318.0421 & $1\times4\times206$ & 128 & 891 & 3540 \\
\noalign{\smallskip}
BD~+08~2211
& 05-Jan-24 & ESPaDOnS & 5315.1807 & $1\times4\times220$ & 123 & 921 & 2903 \\
\noalign{\smallskip}
\hline
\end{tabular}

\tablefoot{
Columns list the target identifiers, observation date, instrument, mid-HJD (Heliocentric Julian Date), exposure time (number of co-added sequences of four sub-exposures with the sub-exposure duration in seconds),   S/N per spectral pixel of Stokes~$I$ for spectral order 45 (503 nm), and mean S/N of the LSD Stokes~$I$ and $V$ profiles.}

\end{table*}

\section{Integration ranges for $B_l$ calculations}

\begin{figure*}[h]
\caption{Zoomed LSD Stokes~$V$ (top panel, solid line) and LSD Stokes~$I$ (bottom panel, solid line) profiles for the four magnetic stars. The vertical dotted grey line indicates the centroid of the line profile and the dashed red lines our integration limits for the calculation of $B_l$. Sub-figure (a) includes in dashed blue lines the choice of integration limits for this target of \citealias{Thomson-Paressant2024}{TP24}.}
\centering
\includegraphics[width=18cm]{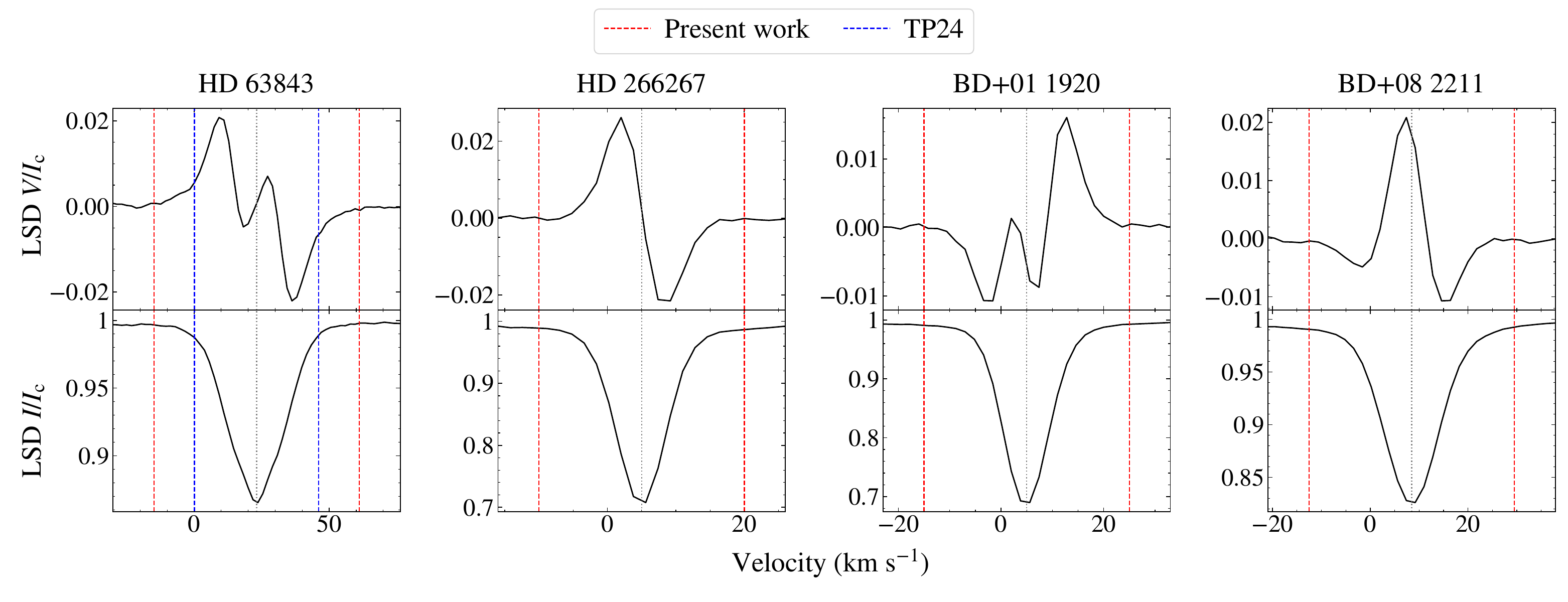}
\label{fig:Integration_limits}
\tablefoot{Since only a single observation is available for each target, the integration limits were determined by inspection, made possible by the pronounced Zeeman signatures observed in the LSD Stokes~$V$ profiles of the DDs. For all targets, we verified that adopting integration ranges up to 1.25 times wider than those selected here yields longitudinal field strengths $B_l$ consistent within uncertainties with the reported values, albeit with larger error bars. This supports the robustness of the chosen integration limits.}

\end{figure*}

\newpage

\section{Parameters and plots of the LSD Stokes~$I$ profile fits}

\begin{table*}[h]
\captionsetup{justification=raggedright,singlelinecheck=false}
\caption{Adopted fit parameters for the upper-limit calculations.}
\centering
\begin{tabular}{l c c c c c c c c}
\hline
\noalign{\smallskip}
Target & Date &
Fit $v$ range (km\,s$^{-1}$) &
Fit $v\sin i$ (km\,s$^{-1}$) &
$v_r$ &
$\sigma_1$ &
$d_1$ &
$\sigma_2$ &
$d_2$ \\
\noalign{\smallskip}
\hline\hline
\noalign{\smallskip}

HD~65900 & 04-Jan-24 &
\multirow{2}{*}{$[10, 90]$} &
\multirow{2}{*}{36} &
47.80 & 5.60 & 1.92 & 5.60 & 1.06 \\

& 08-Jan-24 &
 &  &
48.09 & 5.33 & 1.99 & 5.33 & 0.99 \\

\noalign{\smallskip}
\cdashline{1-9}[0.5pt/2pt]
\noalign{\smallskip}

HD~154228 (33 Oph) & 07-Apr-25 &
$[-200.0, 201.4]$ &
42 &
-29.96 & 202.34 & 0.219 & 5.54 & 2.62 \\

\noalign{\smallskip}
\cdashline{1-9}[0.5pt/2pt]
\noalign{\smallskip}

HD~158716 & 07-Apr-25 &
\multirow{2}{*}{$[-49, 0]$} &
5 &
-24.87 & 7.02 & 1.59 & 0.010 & 0.49 \\

& 16-Jul-14 &
 & 4 &
-25.16 & 6.87 & 1.75 & 0.003 & 0.24 \\

\hline
\end{tabular}

\tablefoot{
All calculations assume a linear limb darkening coefficient $u = 0.5165$ taken from \cite{Claret2019}.
Listed are the star name, date of observation, the velocity range and $v \sin{i}$ employed for each fit, as well as the best-fitting double-Gaussian parameters for each target and observation.}

\label{tab:I_fit_parameters}
\end{table*}

\begin{figure*}[h]
    \centering

    \hspace*{\fill}
    \begin{subfigure}{0.32\textwidth}
        \centering
        \includegraphics[width=\linewidth]{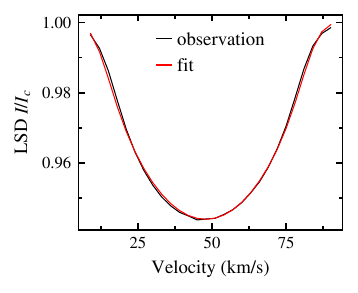}
        \caption{HD~65900 (04-Jan-24, observation 1)}
    \end{subfigure}
    \hspace*{\fill}
    \begin{subfigure}{0.32\textwidth}
        \centering
        \includegraphics[width=\linewidth]{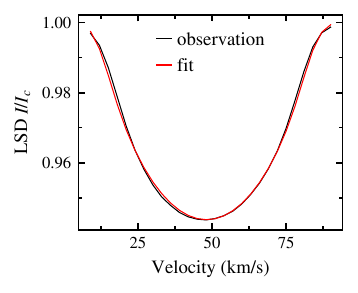}
        \caption{HD~65900 (08-Jan-24)}
    \end{subfigure}
    \hspace*{\fill}
    
    \medskip
    
    \begin{subfigure}{0.32\textwidth}
        \centering
        \includegraphics[width=\linewidth]{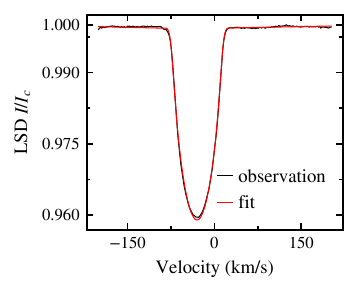}
        \caption{HD~154228 (07-Apr-25)}
    \end{subfigure}
    \hfill
    \begin{subfigure}{0.32\textwidth}
        \centering
        \includegraphics[width=\linewidth]{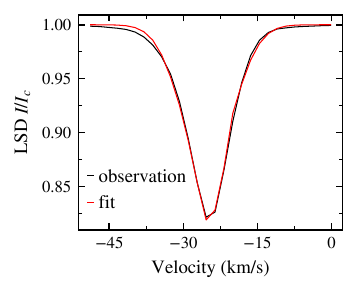}
        \caption{HD~158716 (07-Apr-25)}
    \end{subfigure}
    \hfill
    \begin{subfigure}{0.32\textwidth}
        \centering
        \includegraphics[width=\linewidth]{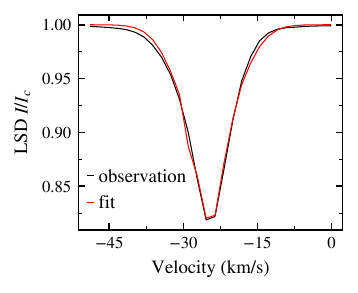}
        \caption{HD~158716 (16-Jul-14, Narval)}
    \end{subfigure}
    
    \medskip
    
    \caption{LSD Stokes~$I$ profile for the different observations of the stars with no field detection (black lines), with the corresponding double-Gaussian fits (red lines). The main parameters for the fits are listed in Table \ref{tab:I_fit_parameters}}
    \label{fig:I_fit}
\end{figure*}

\end{appendix}

\end{document}